\documentstyle[aps]{revtex}
\begin{document}
\centerline{\Large\bf Integrabilities of the $t-J$ Model with Impurities} 
\bigskip
\centerline{\bf Zhan-Ning Hu$^{a,b,}$\footnote{{\bf E-mail:
huzn@aphy01.iphy.ac.cn}} and Fu-Cho Pu$^{b,c,}$\footnote{{\bf E-mail:
pufc@aphy02.iphy.ac.cn}}} \smallskip
\centerline{$^a$CCAST(World Lab.), P.O.Box 8730, Beijing 100080} \centerline{
$^b$Institute of Physics, Academia Sinica, Beijing 100080, China\footnote{
{\bf mail address}}} \centerline{$^c$Department of Physics, Guangzhou
teacher colleague, Guangzhou 510400, China} \centerline{\bf Yupeng Wang
\footnote{{\bf E-mail: ypwang@cl.cryo.ac.cn}}} \centerline{Cryogentic
Laboratory, Academia Sinica, Beijing 100080, China}

\begin{center}
\begin{minipage}{5in}
\centerline{\large\bf   Abstract}
The hamiltonian with magnetic impurities
coupled to the strongly correlated electron
system is constructed from $t-J$ model. And it
is diagonalized exactly  by using the Bethe
ansatz method. Our boundary matrices depend on
the spins of the  electrons. The Kondo
problem in this system is discussed in details.
The  integral equations are derived
 with complex rapidities which describe
the bound states in the system.
The finite-size
corrections for the ground-state energies are obtained.

\smallskip

{\it PACS}: 0550; 7127; 7420                     

\end{minipage}
\end{center}

\newpage

\section{Introduction}

The Kondo problem devoted to study the effect due to the exchange
interaction between the impurity spin and the electron gas has played an
important role in condensed matter physics since its discovery \cite{kon}
.The original treatments in Kondo problem the electron-electron interaction
is discarded. This is reasonable in 3D where the interacting electron system
can be described by Fermi liquid. Recently, much attention has been paid to
the theory of the magnetic impurities in the Fermi liquid and Luttinger
liquid,\cite{222},\cite{333}where the central scheme is the few impurity
coupled with strongly-correlated electron system. Apart from the fundamental
theoretical interests, it is remarkable that the physics implied here can be
accessible experimentally. The recent advances in semiconductor technology
enable to fabricate very narrow quantum wire which can be considered
one-dimensional. and furnishes a real system of Luttinger liquid. Also edge
states in a 2D electron gas for fractional quantum Hall effect can be
considered as Luttinger liquid\cite{Wen}. Intense efforts and much progress
has been made around the subjects from different approaches. Using
bosonization and renormalization techniques, Kane and Fisher\cite{KF}studied
transport of a 1D interacting electron due to potential barriers. Their
results triggered the study of the problem of local perturbations to
Luttinger liquid and Kondo problem in Luttinger liquid. The Kondo problem in
Luttinger liquid was considered by Lee and Toner\cite{LT}. They also
performed the renormalization group calculation and found the crossover of
the Kondo temperature from power law dependence on the Kondo coupling
constant to an exponential one. Relying on poor man's scaling method,
Frusaki and Nagaosa\cite{FG}showed that the Kondo coupling flows to the
strong-coupling regime not only for the antiferromagnetic case but also for
the ferromagnetic case. The boundary conformal field theory\cite{FJ}allows a
classification of critical behavior for Luttinger liquid coupled to a
magnetic impurity. It turns out that there are two possibilities, a local
Fermi liquid with standard low-temperature thermodynamics or a non-Fermi
liquid\cite{FG}. The non-Fermi liquid behavior is induced by the tunneling
effect of conduction electrons through the impurity which depends only on
the bulk properties but not on the details of the impurity\cite{WVP}.
Density matrix renormalization group calculation also supports the same
conclusion\cite{Wx} . In addition the renormalization group flow diagram for
parameters characterizing impurity is more complex and contains fixed points
responsible for the low temperature behaviors when the potential of impurity
is not strong\cite{OF}.

Despite all important progress hitherto made, the problem of few impurities
embedded in a strongly correlated 1D electron system is still far from a
complete understanding. We think that exact solutions of some integrable
models on the subjects are useful from which one can expect to draw definite
conclusions. Indeed Bed\"{u}fig $et$ $al$ has thoroughly solved an
integrable model with impurity coupled with $t-J$ chain\cite{bef}. They
introduce the impurity through a local vertices as in\cite{AJ}. The model
introduced suffers the lack of backward scattering and the presence of
redundant terms in the hamiltonian . Based on Kane and Fisher's observation 
\cite{KF}, we see it is advantageous to use open boundary problem with the
impurities at open ends to study the problem of impurities coupled with
strongly-correlated electron system. The program has been initiated for $
\delta -$ interacting fermi system in \cite{WV} for $t-J$ model in\cite{WDHP}
and for Hubbard model in \cite{WP}.

The $t-J$ model, is considered as one of the most fundamental models in
strongly correlated electron system for its possible relevance for purely
electronic mechanisms for high-$T_c$ superconductivity and heavy-fermion
system. This model is obtained from the Hubbard model as an effective
hamiltonian for the low-energy states in the strong- correlation limit. In
this limit double occupancy of fermions is forbidden, leading to only three
possible states at each lattice site for half spin. Currently, there is
upsurge for its study. Very recently, the Luttinger liquid properties of the 
$t-J$ model are discussed in Ref. \cite{jks}. By solving the functional
relations, the finite-size corrections related to $t-J$ model are dealt with
for the open boundary conditions in Ref. \cite{zhoubat}.The effects about an
integrable impurity coupling to both spin and charge degrees of freedom are
studied in a periodic $t-J$ chain\cite{bef}which we have mentioned above.
The another generalization of the $t-J$ model is given in Ref. \cite{zhang}
by using the one-parametric family of four-dimensional representations of $
gl(2|1)$. It is also a kind of generalization of extended Hubbard model \cite
{kore}{\bf . }

In this paper we expand the study of the Kondo problem in 1D $t-J$ model\cite
{WDHP} by exact solution of open boundary Bethe ansatz. For this purpose we
put two magnetic impurities in both sides of the open $t-J$ model which is a
typical situation for the one-dimensional systems with impurities. The
coupling constants of the impurities with conduction electrons cover from
negative infinity to positive infinity, which means that both the
ferromagnetic Kondo effect and antiferromagnetic Kondo effect can be dealt
with on the same setting. We then construct the hamiltonian for the system
with magnetic impurities from $t-J$ model. The integrability of this model
ensures that both the Yang-Baxter equation and the reflecting Yang-Baxter
equation are satisfied. By using the algebraic Bethe ansatz scheme for open
boundary\cite{Sklyanin} we diagonalize the hamiltonian for the present
system and obtain the Bethe ansatz equations. From which we derive the
nonlinear integral equations governing the thermodynamic properties of the
model for large system. The finite-size corrections for energy of
ground-states in all cases can be calculated.

The arrangement of the present article is as follows. In section 2 the
constructed hamiltonian and its first quantization form are given
explicitly. In section 3 the boundary matrix depending on the rapidity and
spin of the particle is given and all possible integrable cases for the
model are exhausted. The Bethe ansatz equations of the systems for all
integrable cases are derived in section 4. The properties of the ground
state for the cases other than that in \cite{WDHP}are discussed in great
detail in section 5. In the final section the finite-size corrections of the
ground-state energies for chosen cases are obtained.

\section{The Hamiltonian of the Model}

Consider one-dimensional lattice with $G$ sites , $N$ electrons and two
magnetic impurities at both ends. Due to a large on-site Coulomb repulsion
there are at most one particle at one site. The dynamics of the system
governed by a hamiltonian which we construct from the $t-J$ model [22-27].
The conduction electrons can hop ( $t$ ) between the neighbor sites. There
are four types interactions in the model. A spin exchange interaction $J$
and a charge interaction independent of the spin of strength $V$ between the
neighbor conduction electron; Kondo coupling $J_a,J_b$and impurity potential
interactions $V_a,V_b$between the electron and impurities. The hamiltonian
of the system has the form: 
\begin{eqnarray}
H &=&-t\sum_{j=1}^{G-1}\sum_{\sigma =\uparrow \downarrow }(C_{j\sigma
}^{+}C_{j+1\sigma }+C_{j+1\sigma }^{+}\ C_{j\sigma })+J\sum_{j=1}^{G-1}{\bf 
S }_j\cdot {\bf S}_{j+1}+V\sum_{j=1}^{G-1}n_jn_{j+1}  \nonumber \\
&&+J_a{\bf S}_1\cdot {\bf S}_a+V_an_1+J_b{\bf S}_G\cdot {\bf S}_b+V_bn_G,
\label{e1}
\end{eqnarray}
where $C_{j\sigma }^{+}(C_{j\sigma })$ is the creation (annihilation)
operator of the conduction electron with spin $\sigma $ on the site $j$; $
J_{a,b},V_{a,b}$ are the Kondo coupling constants and the impurity
potentials, respectively; ${\bf S}_j=\frac 12\sum_{\sigma ,\sigma ^{\prime
}}C_{j\sigma }^{+}\sigma _{\sigma ,\sigma ^{\prime }}C_{j\sigma ^{\prime }}$
is the spin operator of the conduction electron; $n_j=C_{j\uparrow
}^{+}C_{j\uparrow }+C_{j\downarrow }^{+}C_{j\downarrow }$is the number
operator of the conduction electron; $G$ is the length ( or site number ) of
the system. Some properties of the ground state for $t=1,$ $J=2,$ $V=3/2$
have been reported in Ref. \cite{WDHP}. Following Schultz's notation\cite
{Schultz}we write the translation operators $T_j^{\pm }$: 
\[
T_j^{\pm }\Psi (x_1,\cdots ,x_j,\cdots ,x_N)=\Psi (x_1,\cdots ,x_j\pm
1,\cdots ,x_N), 
\]
where $\Psi (x_1,\cdots ,x_j,\cdots ,x_N)$ is the wave function of $N$
conduction electrons. In first quantization and in appropriate energy units
( $t=1$) the hamiltonian of this system can be written down as 
\begin{equation}
H=-\sum_{j=1}^N\left( T_j^{+}+T_j^{-}\right) +\sum_{j=1}^N\left(
K_{aj}\delta _{x_j,1}+K_{bj}\delta _{x_j,G}+K_j\right)
\end{equation}
where the couplings are denoted by operators $K_{aj}=V_a-\frac{J_a}4+\frac{
J_a}2P_{aj}$ and $K_{bj}=V_b-\frac{J_b}4+\frac{J_b}2P_{bj}$ with the
permutation operators $P_{a(b),j}$ between the spins of the conduction
electron $j$ and the impurities $a,($ $b)$. The operator $K_j$ acts on the
wave function $\Psi $ as 
\[
K_j\Psi (x_1,x_2,\cdots ,x_N)=\sum_{i=1}^N\delta _{x_j,x_i+1}K_{ij}\Psi
(x_1,x_2,\cdots ,x_N), 
\]
where $K_{ij}=V-\frac J4+\frac J2P_{ij}$ describes the interactions between
the conduction electrons with the permutation operator $P_{ij}$ permuting $i$
-th and $j-$ th electron in spin space. We will diagonalize the above
hamiltonian in the following section.

\section{Integrability Conditions}

We write the wave function in region $0\leq x_{Q1}\leq x_{Q2}\leq \cdots
\leq x_{QN}\leq L-1$ as 
\begin{eqnarray}
&&\Psi _{\sigma _1,\sigma _2,\cdots ,\sigma _N}(x_1,x_2,\cdots ,x_N) 
\nonumber \\
&=&\sum_P\sum_{r_1,r_2,\cdots r_N=\pm 1}\varepsilon _P\varepsilon
_rA_{\sigma _{Q1},\sigma _{Q2},\cdots ,\sigma
_{QN}}(r_{PQ1}k_{PQ1},r_{PQ2}k_{PQ2},\cdots ,r_{PQN}k_{PQN})  \nonumber \\
&&\cdot \exp [i\sum_{j=1}^Nr_{Pj}k_{Pj}x_j]
\end{eqnarray}
where the coefficients $A_{\sigma _{Q1},\sigma _{Q2},\cdots ,\sigma
_{QN}}(r_{PQ1}k_{PQ1},r_{PQ2}k_{PQ2},\cdots ,r_{PQN}k_{PQN})$ are also
dependent on the spins of magnetic impurities which are suppressed for
brevity,and $\varepsilon _P=1(-1),$when the parity of $P$ is even(odd)$
,\varepsilon _r=\prod_{j=1}^Nr$ in which $r$ takes the value$+1$or $-1.$ The
boundary $R$ matrix satisfies the reflecting Yang-Baxter equation: 
\begin{equation}
S_{12}(\lambda ,\mu )\stackrel{1}{R}(\lambda )S_{12}(\lambda ,-\mu ) 
\stackrel{2}{R}(\mu )=\stackrel{2}{R}(\mu )S_{12}(\lambda ,-\mu )\stackrel{1 
}{R}(\lambda )S_{12}(\lambda ,\mu ),  \label{byb}
\end{equation}
where operators $\stackrel{1}{R}(\lambda )$ and $\stackrel{2}{R}(\mu )$ are
defined as 
\[
\stackrel{1}{R}(\lambda )=R(\lambda )\otimes id_{V_2},\,\qquad \stackrel{2}{
R }(\mu )=id_{V_1}\otimes R(\mu ) 
\]
for matrix $R\in End(V).$ $S$ matrix satisfies the normal factorizable
condition: 
\begin{equation}
S_{12}(k,\lambda )S_{13}(k,\mu )S_{23}(\lambda ,\mu )=S_{23}(\lambda ,\mu
)S_{13}(k,\mu )S_{12}(k,\lambda ).
\end{equation}
For convenience we set $t=1$. From reflecting Yang-Baxter equation and the
form for $S$ matrix, we refer that the boundary $R$ matrix should have the
form 
\begin{equation}
R=\exp (i\varphi )\frac{q-iC-iP}{q+iC+iP},
\end{equation}
where $P\;$is the permutation operator, $q=\pm \frac 12\cot \frac k2,$ $\pm 
\frac 12\tan \frac k2$ and $C$ is the arbitrary constant. Putting $
K_{a(b),j} $ $=$ $m+lP$ , we have from eq. (\ref{byb}) that 
\begin{equation}
q\left[ \left( m-1\right) ^2-l^2\right] \tan ^2\frac k2+2l\left(
q^2+C^2-1\right) \tan \frac k2+q\left[ \left( m+1\right) ^2-l^2\right] =0.
\end{equation}
This is the restriction imposed on coupling constants in order that our
model (\ref{e1}) to be integrable. The details are as follows.

$J=2,$ $V=-\frac 12$

In this case we know that the scattering matrix in the bulk can be written
as: 
\begin{equation}
S_{12}(k_1,k_2)=\frac{\frac 12\cot \frac{k_1}2-\frac 12\cot \frac{k_2} 2
-iP_{12}}{\frac 12\cot \frac{k_1}2-\frac 12\cot \frac{k_2}2-i}
\end{equation}
where $P_{12}$ is the permutation operator between two electrons. The
boundary $R$ matrix at the left end of the chain takes the form: 
\begin{equation}
R_a(k_j,\sigma _j)=\exp [i\varphi _a(k_j)]\frac{\frac 12\cot \frac{k_j} 2
-iC_a-iP_{aj}}{\frac 12\cot \frac{k_j}2+iC_a+iP_{aj}}.
\end{equation}
The coupling constants $J_a,$ $V_a$ at the left end of the chain are
expressed in terms, of $C_a$ 
\begin{equation}
J_a=-\frac 8{\left( 2C_a\mp 1\right) \left( 2C_a\pm 3\right) },\qquad V_a= 
\frac{3-4C_a^2}{\left( 2C_a\mp 1\right) \left( 2C_a\pm 3\right) },
\label{ee5}
\end{equation}
and 
\begin{equation}
\exp \left[ i\varphi _a(k_j)\right] =\frac{J_a(\cot \frac{k_j}2+2iC_a)\exp
(ik_j)+i[4+(4V_a-J_a)\exp (ik_j)]}{J_a(\cot \frac{k_j}2-2iC_a)\exp
(-ik_j)-i[4+(4V_a-J_a)\exp (-ik_j)]}.  \label{ee6}
\end{equation}
The boundary $R$ matrix at the right end of the chain has the form: 
\begin{equation}
R_b(-k_j,\sigma _j)=\exp [-2ik_j(G+1)+i\varphi _b(k_j)]\frac{\frac 12\cot 
\frac{k_j}2-iC_b-iP_{bj}}{\frac 12\cot \frac{k_j}2+iC_b+iP_{bj}}.
\end{equation}
Similar relations exist for $J_b,$ $V_b$ and $\varphi _b(k_j)$ by merely
substituting indices $a$ in (\ref{ee5}) and (\ref{ee6}) by $b.$

\subsection{$J=-2$ $V=\frac 12$}

The boundary $R$ matrices have the forms: 
\begin{equation}
R_a(k_j,\sigma _j)=\exp [i\varphi _a(k_j)]\frac{\frac 12\tan \frac{k_j} 2
+iC_a+iP_{aj}}{\frac 12\tan \frac{k_j}2-iC_a-iP_{aj}},
\end{equation}
\begin{equation}
R_b(-k_j,\sigma _j)=\exp [-2ik_j(G+1)+i\varphi _b(k_j)]\frac{\frac 12\tan 
\frac{k_j}2+iC_b+iP_{bj}}{\frac 12\tan \frac{k_j}2-iC_b-iP_{bj}}.
\end{equation}
$\varphi _a(k_j)$ and $\varphi _b(k_j)$ are the same as in the proceeding.
Now the coupling constants should be written in terms of the arbitrary
parameter $C_a$ in the form 
\begin{equation}
J_a=\frac 8{\left( 2C_a\mp 1\right) \left( 2C_a\pm 3\right) },\qquad V_a= 
\frac{4C_a^2-3}{\left( 2C_a\mp 1\right) \left( 2C_a\pm 3\right) }
\end{equation}
$J_{b,}$ $V_b$ have the same expressions except with the substitution of
indices $a$ by $b.$ Correspondingly, the scattering matrix $S$ in the bulk
for two conduction electrons is 
\begin{equation}
S_{12}(k_1,k_2)=\frac{\frac 12\tan \frac{k_1}2-\frac 12\tan \frac{k_2} 2
+iP_{12}}{\frac 12\tan \frac{k_1}2-\frac 12\tan \frac{k_2}2+i}.
\end{equation}

\subsection{$J=2$ $V=\frac 32$}

In this case the dependence of coupling constants on parameter $C_a$takes
the form 
\begin{equation}
J_a=-\frac 8{\left( 2C_a\mp 1\right) \left( 2C_a\pm 3\right) },\qquad V_a= 
\frac{4C_a^2-7}{\left( 2C_a\mp 1\right) \left( 2C_a\pm 3\right) }.
\end{equation}
$J_{b,}$ $V_b$ have the same expressions by the substituting of indices $a$
by $b.$ The scattering matrix in the bulk is 
\begin{equation}
S_{12}(k_1,k_2)=-\frac{\frac 12\tan \frac{k_1}2-\frac 12\tan \frac{k_2} 2
-iP_{12}}{\frac 12\tan \frac{k_1}2-\frac 12\tan \frac{k_2}2+i}.
\end{equation}
The boundary $R$ matrices are 
\begin{equation}
R_a(k_j,\sigma _j)=\exp [i\varphi _a(k_j)]\frac{\frac 12\tan \frac{k_j} 2
-iC_a-iP_{aj}}{\frac 12\tan \frac{k_j}2+iC_a+iP_{aj}},
\end{equation}
\begin{equation}
R_b(-k_j,\sigma _j)=\exp [-2ik_j(G+1)+i\varphi _b(k_j)]\frac{\frac 12\tan 
\frac{k_j}2-iC_b-iP_{bj}}{\frac 12\tan \frac{k_j}2+iC_b+iP_{bj}}.
\end{equation}

\subsection{$J=-2$ $V=-\frac 32$}

The coupling constants have the forms 
\begin{equation}
J_a=\frac 8{\left( 2C_a\mp 1\right) \left( 2C_a\pm 3\right) },\qquad V_a= 
\frac{7-4C_a^2}{\left( 2C_a\mp 1\right) \left( 2C_a\pm 3\right) }.
\end{equation}
$J_{b,}$ $V_b$ have the same expressions by the substituting of indices $a$
by $b.$ The scattering matrix in the bulk is 
\begin{equation}
S_{12}(k_1,k_2)=-\frac{\frac 12\cot \frac{k_1}2-\frac 12\cot \frac{k_2} 2
+iP_{12}}{\frac 12\cot \frac{k_1}2-\frac 12\cot \frac{k_2}2-i}.
\end{equation}
The boundary $R$ matrices are 
\begin{equation}
R_a(k_j,\sigma _j)=\exp [i\varphi _a(k_j)]\frac{\frac 12\cot \frac{k_j} 2
+iC_a+iP_{aj}}{\frac 12\cot \frac{k_j}2-iC_a-iP_{aj}},
\end{equation}
\begin{equation}
R_b(-k_j,\sigma _j)=\exp [-2ik_j(G+1)+i\varphi _b(k_j)]\frac{\frac 12\cot 
\frac{k_j}2+iC_b+iP_{bj}}{\frac 12\cot \frac{k_j}2-iC_b-iP_{bj}},
\end{equation}
The expressions for boundary matrices depending on both the moment of the
particle and the spin of the electron are new. The expressions of $S$ matrix
in the bulk have been obtained before in Ref. \cite{Sch},but they are
different from ours.

\section{Bethe Ansatz Equations}

By using the standard Bethe ansatz procedure, we can diagonalize the
hamiltonian (\ref{e1}) \cite{Sklyanin}and obtain the following Bethe Ansatz
equations. When $J=2$ and $V=-\frac 12$, setting 
\begin{eqnarray}
S_{j0}(k_j,k_0) &=&\frac{\frac 12\cot \frac{k_j}2-\frac 12\cot \frac{k_0} 2
-iP_{j0}}{\frac 12\cot \frac{k_j}2-\frac 12\cot \frac{k_0}2-i}, \\
S_{jN+1}(k_j,k_{N+1}) &=&\frac{\frac 12\cot \frac{k_j}2-\frac 12\cot \frac{
k_{N+1}}2-iP_{jN+1}}{\frac 12\cot \frac{k_j}2-\frac 12\cot \frac{k_{N+1}}2-i}
,
\end{eqnarray}
where $\cot \frac{k_0}2=2iC_a,\cot \frac{k_{N+1}}2=2iC_{b,}$ $P_{j0}\equiv
P_{aj},P_{jN+1}\equiv P_{bj},$ we can write down the boundary $R$ matrices
as the forms: 
\begin{equation}
R_a(k_j,\sigma _j)=\exp \left[ i\varphi _a(k_j)\right] \frac{\frac 12\cot 
\frac{k_j}2-iC_a-i}{\frac 12\cot \frac{k_j}2+iC_a+i}\frac{S_{j0}(k_j,k_0)}{
S_{j0}(-k_j,k_0)},
\end{equation}
\begin{equation}
R_b(-k_j,\sigma _j)=\exp \left[ -2ik_j(G+1)+i\varphi _b(k_j)\right] \frac{
\frac 12\cot \frac{k_j}2-iC_b-i}{\frac 12\cot \frac{k_j}2+iC_b+i}\frac{
S_{jN+1}(k_j,k_{N+1})}{S_{jN+1}(-k_j,k_{N+1})}.
\end{equation}
Define 
\begin{equation}
T(\lambda )=S_{\tau j}(\lambda )S_{\tau 0}(\lambda )S_{\tau 1}(\lambda
)\cdots S_{\tau j-1}(\lambda )S_{\tau j+1}(\lambda )\cdots S_{\tau
N}(\lambda )S_{\tau N+1}(\lambda )
\end{equation}
with 
\begin{equation}
S_{\tau l}(\lambda )=\frac{\lambda -\frac 12\cot \frac{k_l}2-iP_{\tau l}}{
\lambda -\frac 12\cot \frac{k_l}2-i},\qquad l=0,1,\cdots ,N+1.
\end{equation}
We get the equation 
\[
\left. Tr\left[ T(\lambda )T^{-1}(-\lambda )\right] \right| _{\lambda =\frac 
12\cot \frac{k_j}2}\Phi \qquad \qquad \qquad \qquad \qquad \qquad \qquad
\qquad \qquad \qquad \qquad 
\]
\begin{eqnarray}
&=&\frac{2i-\cot \frac{k_j}2}{i-\cot \frac{k_j}2}\frac{\frac 12\cot \frac{
k_j }2+iC_a+i}{\frac 12\cot \frac{k_j}2-iC_a-i}\frac{\frac 12\cot \frac{k_j} 
2+iC_b+i}{\frac 12\cot \frac{k_j}2-iC_b-i}  \nonumber \\
&&\cdot \exp \left[ -i\varphi _a(k_j)-i\varphi _b(k_j)+2ik_j(G+1)\right] \Phi
\end{eqnarray}
where $\Phi $ is the eigenstate of the system. Then the Bethe ansatz
equations can be expressed as 
\[
\exp \left[ 2ik_j(G+1)-i\varphi _a(k_j)-i\varphi _b(k_j)\right] \frac{\frac 1
2\cot \frac{k_j}2+iC_a+i}{\frac 12\cot \frac{k_j}2-iC_a-i}\frac{\frac 12\cot 
\frac{k_j}2+iC_b+i}{\frac 12\cot \frac{k_j}2-iC_b-i}\qquad 
\]
\begin{equation}
\qquad \qquad \qquad \qquad =\prod_{\beta =1}^M\frac{\frac 12\cot \frac{k_j} 
2-\lambda _\beta +\frac i2}{\frac 12\cot \frac{k_j}2-\lambda _\beta -\frac i2
}\frac{\frac 12\cot \frac{k_j}2+\lambda _\beta +\frac i2}{\frac 12\cot \frac{
k_j}2+\lambda _\beta -\frac i2},\qquad (j=1,2,\cdots ,N)
\end{equation}
\[
\frac{(\lambda _\alpha +\frac i2)^2+C_a^2}{(\lambda _\alpha -\frac i2
)^2+C_a^2}\frac{(\lambda _\alpha +\frac i2)^2+C_b^2}{(\lambda _\alpha -\frac 
i2)^2+C_b^2}\prod_{l=1}^N\frac{\lambda _\alpha -\frac 12\cot \frac{ k_l }2+
\frac i2}{\lambda _\alpha -\frac 12\cot \frac{k_l}2-\frac i2}\frac{ \lambda
_\alpha +\frac 12\cot \frac{k_l}2+\frac i2}{\lambda _\alpha +\frac 12\cot 
\frac{k_l}2-\frac i2} 
\]

\begin{equation}
=\prod_{\beta =1(\beta \neq \alpha )}^M\frac{\lambda _\alpha -\lambda _\beta
+i}{\lambda _\alpha -\lambda _\beta -i}\frac{\lambda _\alpha +\lambda _\beta
+i}{\lambda _\alpha +\lambda _\beta -i}\qquad (\alpha =1,2,\cdots ,M).
\end{equation}
$M$ is the number of down spins and $N$ is the number of the electrons. The
function $\varphi $ is denoted by expression ( \ref{ee6}). Similarly, when $
J=-2$ and $V=\frac 12,$ we can write down the Bethe Ansatz equations as the
forms: 
\[
\exp \left[ 2ik_j(G+1)-i\varphi _a(k_j)-i\varphi _b(k_j)\right] \frac{\frac 1
2\tan \frac{k_j}2-iC_a-i}{\frac 12\tan \frac{k_j}2+iC_a+i}\frac{\frac 12\tan 
\frac{k_j}2-iC_b-i}{\frac 12\tan \frac{k_j}2+iC_b+i}\qquad 
\]
\begin{equation}
\qquad \qquad \qquad \qquad =\prod_{\beta =1}^M\frac{\frac 12\tan \frac{k_j} 
2-\lambda _\beta -\frac i2}{\frac 12\tan \frac{k_j}2-\lambda _\beta +\frac i2
}\frac{\frac 12\tan \frac{k_j}2+\lambda _\beta -\frac i2}{\frac 12\tan \frac{
k_j}2+\lambda _\beta +\frac i2},\qquad (j=1,2,\cdots ,N)
\end{equation}
\[
\frac{(\lambda _\alpha +\frac i2)^2+C_a^2}{(\lambda _\alpha -\frac i2
)^2+C_a^2}\frac{(\lambda _\alpha +\frac i2)^2+C_b^2}{(\lambda _\alpha -\frac 
i2)^2+C_b^2}\prod_{l=1}^N\frac{\lambda _\alpha -\frac 12\tan \frac{ k_l }2+
\frac i2}{\lambda _\alpha -\frac 12\tan \frac{k_l}2-\frac i2}\frac{ \lambda
_\alpha +\frac 12\tan \frac{k_l}2+\frac i2}{\lambda _\alpha +\frac 12\tan 
\frac{k_l}2-\frac i2} 
\]

\begin{equation}
=\prod_{\beta =1(\beta \neq \alpha )}^M\frac{\lambda _\alpha -\lambda _\beta
+i}{\lambda _\alpha -\lambda _\beta -i}\frac{\lambda _\alpha +\lambda _\beta
+i}{\lambda _\alpha +\lambda _\beta -i}\qquad (\alpha =1,2,\cdots ,M).
\end{equation}
When $J=2$ and $V=\frac 32,$ we have that 
\begin{eqnarray*}
&&\exp \left[ 2ik_j(G+1)-i\varphi _a(k_j)-i\varphi _b(k_j)\right] \frac{
\frac 12\tan \frac{k_j}2+iC_a+i}{\frac 12\tan \frac{k_j}2-iC_a-i}\frac{\frac 
12\tan \frac{k_j}2+iC_b+i}{\frac 12\tan \frac{k_j}2-iC_b-i} \\
&&\cdot \prod_{l=1(l\neq j)}^N\frac{\frac 12\tan \frac{k_j}2-\frac 12\tan 
\frac{k_l}2+i}{\frac 12\tan \frac{k_j}2-\frac 12\tan \frac{k_l}2-i}\frac{
\frac 12\tan \frac{k_j}2+\frac 12\tan \frac{k_l}2+i}{\frac 12\tan \frac{k_j} 
2+\frac 12\tan \frac{k_l}2-i}
\end{eqnarray*}
\begin{equation}
\qquad \qquad =\prod_{\beta =1}^M\frac{\frac 12\tan \frac{k_j}2-\lambda
_\beta +\frac i2}{\frac 12\tan \frac{k_j}2-\lambda _\beta -\frac i2}\frac{
\frac 12\tan \frac{k_j}2+\lambda _\beta +\frac i2}{\frac 12\tan \frac{k_j} 2
+\lambda _\beta -\frac i2},\qquad (j=1,2,\cdots ,N),
\end{equation}
\[
\frac{(\lambda _\alpha +\frac i2)^2+C_a^2}{(\lambda _\alpha -\frac i2
)^2+C_a^2}\frac{(\lambda _\alpha +\frac i2)^2+C_b^2}{(\lambda _\alpha -\frac 
i2)^2+C_b^2}\prod_{l=1}^N\frac{\lambda _\alpha -\frac 12\tan \frac{ k_l }2+
\frac i2}{\lambda _\alpha -\frac 12\tan \frac{k_l}2-\frac i2}\frac{ \lambda
_\alpha +\frac 12\tan \frac{k_l}2+\frac i2}{\lambda _\alpha +\frac 12\tan 
\frac{k_l}2-\frac i2} 
\]

\begin{equation}
=\prod_{\beta =1(\beta \neq \alpha )}^M\frac{\lambda _\alpha -\lambda _\beta
+i}{\lambda _\alpha -\lambda _\beta -i}\frac{\lambda _\alpha +\lambda _\beta
+i}{\lambda _\alpha +\lambda _\beta -i}\qquad (\alpha =1,2,\cdots ,M).
\end{equation}
When $J=-2$ and $V=-\frac 32,$ we get that 
\begin{eqnarray*}
&&\exp \left[ 2ik_j(G+1)-i\varphi _a(k_j)-i\varphi _b(k_j)\right] \frac{
\frac 12\cot \frac{k_j}2-iC_a-i}{\frac 12\cot \frac{k_j}2+iC_a+i}\frac{\frac 
12\cot \frac{k_j}2-iC_b-i}{\frac 12\cot \frac{k_j}2+iC_b+i} \\
&&\cdot \prod_{l=1(l\neq j)}^N\frac{\frac 12\cot \frac{k_j}2-\frac 12\cot 
\frac{k_l}2-i}{\frac 12\cot \frac{k_j}2-\frac 12\cot \frac{k_l}2+i}\frac{
\frac 12\cot \frac{k_j}2+\frac 12\cot \frac{k_l}2-i}{\frac 12\cot \frac{k_j} 
2+\frac 12\cot \frac{k_l}2+i}
\end{eqnarray*}
\begin{equation}
\qquad \qquad =\prod_{\beta =1}^M\frac{\frac 12\cot \frac{k_j}2-\lambda
_\beta -\frac i2}{\frac 12\cot \frac{k_j}2-\lambda _\beta +\frac i2}\frac{
\frac 12\cot \frac{k_j}2+\lambda _\beta -\frac i2}{\frac 12\cot \frac{k_j} 2
+\lambda _\beta +\frac i2},\qquad (j=1,2,\cdots ,N),
\end{equation}
\[
\frac{(\lambda _\alpha +\frac i2)^2+C_a^2}{(\lambda _\alpha -\frac i2
)^2+C_a^2}\frac{(\lambda _\alpha +\frac i2)^2+C_b^2}{(\lambda _\alpha -\frac 
i2)^2+C_b^2}\prod_{l=1}^N\frac{\lambda _\alpha -\frac 12\cot \frac{ k_l }2+
\frac i2}{\lambda _\alpha -\frac 12\cot \frac{k_l}2-\frac i2}\frac{ \lambda
_\alpha +\frac 12\cot \frac{k_l}2+\frac i2}{\lambda _\alpha +\frac 12\cot 
\frac{k_l}2-\frac i2} 
\]

\begin{equation}
=\prod_{\beta =1(\beta \neq \alpha )}^M\frac{\lambda _\alpha -\lambda _\beta
+i}{\lambda _\alpha -\lambda _\beta -i}\frac{\lambda _\alpha +\lambda _\beta
+i}{\lambda _\alpha +\lambda _\beta -i}\qquad (\alpha =1,2,\cdots ,M).
\end{equation}
Here the function $\varphi _a(k_{j)}$ is expressed by equation (\ref{ee6})
and $\varphi _b(k_j)$ has the same expression as relation (\ref{ee6}) with
the substitution of index $a$ by $b.$ $M$ is the number of down spins and $N$
is the number of the electrons. It should be noted that in the above Bethe
Ansatz equations we have choose the boundary $R$ matrices as the form as the
ones in Section 3, where the boundary matrices depend on the spin parameter.
If the $R$ matrix is dependent on the spin of the electron only at one end
of the chain, for example, we denote by $R_b(k_j,\widehat{\sigma }_j)$ the
boundary matrix at right end of the chain independent on the spin $\sigma _j$
. The Bethe Ansatz equations for $J=2,$ $V=-\frac 12$ take the form: 
\begin{equation}
\frac{\exp \left[ -i\varphi _a(k_j)\right] }{R_b(-k_j,\widehat{\sigma }_j)} 
\frac{\frac 12\cot \frac{k_j}2+iC_a+i}{\frac 12\cot \frac{k_j}2-iC_a-i}
=\prod_{\beta =1}^M\frac{\frac 12\cot \frac{k_j}2-\lambda _\beta +\frac i2}{
\frac 12\cot \frac{k_j}2-\lambda _\beta -\frac i2}\frac{\frac 12\cot \frac{
k_j}2+\lambda _\beta +\frac i2}{\frac 12\cot \frac{k_j}2+\lambda _\beta -
\frac i2},
\end{equation}
\[
\qquad \qquad \qquad \qquad \qquad \qquad (j=1,2,\cdots ,N) 
\]
\[
\frac{(\lambda _\alpha +\frac i2)^2+C_a^2}{(\lambda _\alpha -\frac i2
)^2+C_a^2}\prod_{l=1}^N\frac{\lambda _\alpha -\frac 12\cot \frac{k_l} 2+
\frac i2}{\lambda _\alpha -\frac 12\cot \frac{k_l}2-\frac i2}\frac{\lambda
_\alpha +\frac 12\cot \frac{k_l}2+\frac i2}{\lambda _\alpha +\frac 12\cot 
\frac{k_l}2-\frac i2} 
\]

\begin{equation}
=\prod_{\beta =1(\beta \neq \alpha )}^M\frac{\lambda _\alpha -\lambda _\beta
+i}{\lambda _\alpha -\lambda _\beta -i}\frac{\lambda _\alpha +\lambda _\beta
+i}{\lambda _\alpha +\lambda _\beta -i},\qquad (\alpha =1,2,\cdots ,M)
\end{equation}
Similarly, when the boundary matrix at the left end of the spin is
independent on the spin of the electron, denoted by $R_a(k_j,\widehat{\sigma 
}_j),$ we have that 
\[
\frac{\exp \left[ 2ik_j(G+1)-i\varphi _b(k_j)\right] }{R_a(k_j,\widehat{
\sigma }_j)}\frac{\frac 12\cot \frac{k_j}2+iC_b+i}{\frac 12\cot \frac{k_j} 2
-iC_b-i}\qquad \qquad \qquad \qquad \qquad 
\]
\begin{equation}
\qquad \qquad \qquad =\prod_{\beta =1}^M\frac{\frac 12\cot \frac{k_j} 2
-\lambda _\beta +\frac i2}{\frac 12\cot \frac{k_j}2-\lambda _\beta -\frac i2 
}\frac{\frac 12\cot \frac{k_j}2+\lambda _\beta +\frac i2}{\frac 12\cot \frac{
k_j}2+\lambda _\beta -\frac i2},\qquad (j=1,2,\cdots ,N)
\end{equation}
\[
\frac{(\lambda _\alpha +\frac i2)^2+C_a^2}{(\lambda _\alpha -\frac i2
)^2+C_a^2}\prod_{l=1}^N\frac{\lambda _\alpha -\frac 12\cot \frac{k_l} 2+
\frac i2}{\lambda _\alpha -\frac 12\cot \frac{k_l}2-\frac i2}\frac{\lambda
_\alpha +\frac 12\cot \frac{k_l}2+\frac i2}{\lambda _\alpha +\frac 12\cot 
\frac{k_l}2-\frac i2} 
\]

\begin{equation}
=\prod_{\beta =1(\beta \neq \alpha )}^M\frac{\lambda _\alpha -\lambda _\beta
+i}{\lambda _\alpha -\lambda _\beta -i}\frac{\lambda _\alpha +\lambda _\beta
+i}{\lambda _\alpha +\lambda _\beta -i},\qquad (\alpha =1,2,\cdots ,M)
\end{equation}
where the number of down spins should less than $N+2$ and $N$ is the number
of the conduction electrons in the system. Furthermore, we get that 
\begin{equation}
\frac{\exp \left[ -i\varphi _a(k_j)\right] }{R_b(-k_j,\widehat{\sigma }_j)} 
\frac{\frac 12\tan \frac{k_j}2-iC_a-i}{\frac 12\tan \frac{k_j}2+iC_a+i}
=\prod_{\beta =1}^M\frac{\frac 12\tan \frac{k_j}2-\lambda _\beta -\frac i2}{
\frac 12\tan \frac{k_j}2-\lambda _\beta +\frac i2}\frac{\frac 12\tan \frac{
k_j}2+\lambda _\beta -\frac i2}{\frac 12\tan \frac{k_j}2+\lambda _\beta +
\frac i2},
\end{equation}
\[
\qquad \qquad \qquad \qquad \qquad (j=1,2,\cdots ,N) 
\]
\[
\frac{(\lambda _\alpha +\frac i2)^2+C_a^2}{(\lambda _\alpha -\frac i2
)^2+C_a^2}\prod_{l=1}^N\frac{\lambda _\alpha -\frac 12\tan \frac{k_l} 2+
\frac i2}{\lambda _\alpha -\frac 12\tan \frac{k_l}2-\frac i2}\frac{\lambda
_\alpha +\frac 12\tan \frac{k_l}2+\frac i2}{\lambda _\alpha +\frac 12\tan 
\frac{k_l}2-\frac i2} 
\]

\begin{equation}
\qquad \qquad =\prod_{\beta =1(\beta \neq \alpha )}^M\frac{\lambda _\alpha
-\lambda _\beta +i}{\lambda _\alpha -\lambda _\beta -i}\frac{\lambda _\alpha
+\lambda _\beta +i}{\lambda _\alpha +\lambda _\beta -i}\qquad (\alpha
=1,2,\cdots ,M)
\end{equation}
and 
\[
\frac{\exp \left[ 2ik_j(G+1)-i\varphi _b(k_j)\right] }{R_a(k_j,\widehat{
\sigma }_j)}\frac{\frac 12\tan \frac{k_j}2-iC_b-i}{\frac 12\tan \frac{k_j} 2
+iC_b+i}\qquad \qquad \qquad \qquad \qquad 
\]
\begin{equation}
\qquad =\prod_{\beta =1}^M\frac{\frac 12\tan \frac{k_j}2-\lambda _\beta -
\frac i2}{\frac 12\tan \frac{k_j}2-\lambda _\beta +\frac i2}\frac{\frac 12
\tan \frac{k_j}2+\lambda _\beta -\frac i2}{\frac 12\tan \frac{k_j} 2+\lambda
_\beta +\frac i2},\qquad (j=1,2,\cdots ,N)
\end{equation}
\[
\frac{(\lambda _\alpha +\frac i2)^2+C_b^2}{(\lambda _\alpha -\frac i2
)^2+C_b^2}\prod_{l=1}^N\frac{\lambda _\alpha -\frac 12\tan \frac{k_l} 2+
\frac i2}{\lambda _\alpha -\frac 12\tan \frac{k_l}2-\frac i2}\frac{\lambda
_\alpha +\frac 12\tan \frac{k_l}2+\frac i2}{\lambda _\alpha +\frac 12\tan 
\frac{k_l}2-\frac i2} 
\]

\begin{equation}
=\prod_{\beta =1(\beta \neq \alpha )}^M\frac{\lambda _\alpha -\lambda _\beta
+i}{\lambda _\alpha -\lambda _\beta -i}\frac{\lambda _\alpha +\lambda _\beta
+i}{\lambda _\alpha +\lambda _\beta -i}\qquad (\alpha =1,2,\cdots ,M)
\end{equation}
for the case of $J=-2,$ $V=\frac 12.$ 
\[
\frac{\exp \left[ -i\varphi _a(k_j)\right] }{R_b(-k_j,\widehat{\sigma }_j)} 
\frac{\frac 12\tan \frac{k_j}2+iC_a+i}{\frac 12\tan \frac{k_j}2-iC_a-i}
\prod_{l=1(l\neq j)}^N\frac{\frac 12\tan \frac{k_j}2-\frac 12\tan \frac{k_l} 
2+i}{\frac 12\tan \frac{k_j}2-\frac 12\tan \frac{k_l}2-i}\frac{\frac 12\tan 
\frac{k_j}2+\frac 12\tan \frac{k_l}2+i}{\frac 12\tan \frac{k_j}2+\frac 12
\tan \frac{k_l}2-i} 
\]
\begin{equation}
\qquad \qquad =\prod_{\beta =1}^M\frac{\frac 12\tan \frac{k_j}2-\lambda
_\beta +\frac i2}{\frac 12\tan \frac{k_j}2-\lambda _\beta -\frac i2}\frac{
\frac 12\tan \frac{k_j}2+\lambda _\beta +\frac i2}{\frac 12\tan \frac{k_j} 2
+\lambda _\beta -\frac i2},\qquad (j=1,2,\cdots ,N),
\end{equation}
\[
\frac{(\lambda _\alpha +\frac i2)^2+C_a^2}{(\lambda _\alpha -\frac i2
)^2+C_a^2}\prod_{l=1}^N\frac{\lambda _\alpha -\frac 12\tan \frac{k_l} 2+
\frac i2}{\lambda _\alpha -\frac 12\tan \frac{k_l}2-\frac i2}\frac{\lambda
_\alpha +\frac 12\tan \frac{k_l}2+\frac i2}{\lambda _\alpha +\frac 12\tan 
\frac{k_l}2-\frac i2}\qquad \qquad \qquad 
\]

\begin{equation}
\qquad \qquad \qquad =\prod_{\beta =1(\beta \neq \alpha )}^M\frac{\lambda
_\alpha -\lambda _\beta +i}{\lambda _\alpha -\lambda _\beta -i}\frac{\lambda
_\alpha +\lambda _\beta +i}{\lambda _\alpha +\lambda _\beta -i}\qquad
(\alpha =1,2,\cdots ,M)
\end{equation}
and 
\begin{eqnarray*}
&&\frac{\exp \left[ 2ik_j(G+1)-i\varphi _b(k_j)\right] }{R_a(k_j,\widehat{
\sigma }_j)}\frac{\frac 12\tan \frac{k_j}2+iC_b+i}{\frac 12\tan \frac{k_j} 2
-iC_b-i} \\
&&\cdot \prod_{l=1(l\neq j)}^N\frac{\frac 12\tan \frac{k_j}2-\frac 12\tan 
\frac{k_l}2+i}{\frac 12\tan \frac{k_j}2-\frac 12\tan \frac{k_l}2-i}\frac{
\frac 12\tan \frac{k_j}2+\frac 12\tan \frac{k_l}2+i}{\frac 12\tan \frac{k_j} 
2+\frac 12\tan \frac{k_l}2-i}\qquad \qquad \qquad \qquad
\end{eqnarray*}
\begin{equation}
\qquad \qquad \qquad =\prod_{\beta =1}^M\frac{\frac 12\tan \frac{k_j} 2
-\lambda _\beta +\frac i2}{\frac 12\tan \frac{k_j}2-\lambda _\beta -\frac i2 
}\frac{\frac 12\tan \frac{k_j}2+\lambda _\beta +\frac i2}{\frac 12\tan \frac{
k_j}2+\lambda _\beta -\frac i2},\qquad (j=1,2,\cdots ,N),
\end{equation}
\[
\frac{(\lambda _\alpha +\frac i2)^2+C_b^2}{(\lambda _\alpha -\frac i2
)^2+C_b^2}\prod_{l=1}^N\frac{\lambda _\alpha -\frac 12\tan \frac{k_l} 2+
\frac i2}{\lambda _\alpha -\frac 12\tan \frac{k_l}2-\frac i2}\frac{\lambda
_\alpha +\frac 12\tan \frac{k_l}2+\frac i2}{\lambda _\alpha +\frac 12\tan 
\frac{k_l}2-\frac i2}\qquad \qquad 
\]

\begin{equation}
=\prod_{\beta =1(\beta \neq \alpha )}^M\frac{\lambda _\alpha -\lambda _\beta
+i}{\lambda _\alpha -\lambda _\beta -i}\frac{\lambda _\alpha +\lambda _\beta
+i}{\lambda _\alpha +\lambda _\beta -i}\qquad (\alpha =1,2,\cdots ,M)
\end{equation}
for the case of $J=2,$ $V=\frac 32$ when boundary matrix only at one end of
the chain rely on the spin parameter of the electron. Finally, for the case
of $J=-2$ and $V=-\frac 32,$ the Bethe Ansatz equations take the form: 
\[
\frac{\exp \left[ -i\varphi _a(k_j)\right] }{R_b(-k_j,\widehat{\sigma }_j)} 
\frac{\frac 12\cot \frac{k_j}2-iC_a-i}{\frac 12\cot \frac{k_j}2+iC_a+i}
\prod_{l=1(l\neq j)}^N\frac{\frac 12\cot \frac{k_j}2-\frac 12\cot \frac{k_l} 
2-i}{\frac 12\cot \frac{k_j}2-\frac 12\cot \frac{k_l}2+i}\frac{\frac 12\cot 
\frac{k_j}2+\frac 12\cot \frac{k_l}2-i}{\frac 12\cot \frac{k_j}2+\frac 12
\cot \frac{k_l}2+i} 
\]
\begin{equation}
\qquad \qquad \qquad \qquad =\prod_{\beta =1}^M\frac{\frac 12\cot \frac{k_j} 
2-\lambda _\beta -\frac i2}{\frac 12\cot \frac{k_j}2-\lambda _\beta +\frac i2
}\frac{\frac 12\cot \frac{k_j}2+\lambda _\beta -\frac i2}{\frac 12\cot \frac{
k_j}2+\lambda _\beta +\frac i2},\qquad (j=1,2,\cdots ,N),
\end{equation}
\[
\frac{(\lambda _\alpha +\frac i2)^2+C_a^2}{(\lambda _\alpha -\frac i2
)^2+C_a^2}\prod_{l=1}^N\frac{\lambda _\alpha -\frac 12\cot \frac{k_l} 2+
\frac i2}{\lambda _\alpha -\frac 12\cot \frac{k_l}2-\frac i2}\frac{\lambda
_\alpha +\frac 12\cot \frac{k_l}2+\frac i2}{\lambda _\alpha +\frac 12\cot 
\frac{k_l}2-\frac i2} 
\]

\begin{equation}
\qquad \qquad \qquad =\prod_{\beta =1(\beta \neq \alpha )}^M\frac{\lambda
_\alpha -\lambda _\beta +i}{\lambda _\alpha -\lambda _\beta -i}\frac{\lambda
_\alpha +\lambda _\beta +i}{\lambda _\alpha +\lambda _\beta -i}\qquad
(\alpha =1,2,\cdots ,M)
\end{equation}
and 
\begin{eqnarray*}
&&\frac{\exp \left[ 2ik_j(G+1)-i\varphi _b(k_j)\right] }{R_a(k_j,\widehat{
\sigma }_j)}\frac{\frac 12\cot \frac{k_j}2-iC_b-i}{\frac 12\cot \frac{k_j} 2
+iC_b+i}\qquad \qquad \\
&&\cdot \prod_{l=1(l\neq j)}^N\frac{\frac 12\cot \frac{k_j}2-\frac 12\cot 
\frac{k_l}2-i}{\frac 12\cot \frac{k_j}2-\frac 12\cot \frac{k_l}2+i}\frac{
\frac 12\cot \frac{k_j}2+\frac 12\cot \frac{k_l}2-i}{\frac 12\cot \frac{k_j} 
2+\frac 12\cot \frac{k_l}2+i}
\end{eqnarray*}
\begin{equation}
\qquad \qquad \qquad =\prod_{\beta =1}^M\frac{\frac 12\cot \frac{k_j} 2
-\lambda _\beta -\frac i2}{\frac 12\cot \frac{k_j}2-\lambda _\beta +\frac i2 
}\frac{\frac 12\cot \frac{k_j}2+\lambda _\beta -\frac i2}{\frac 12\cot \frac{
k_j}2+\lambda _\beta +\frac i2},\qquad (j=1,2,\cdots ,N),
\end{equation}
\[
\frac{(\lambda _\alpha +\frac i2)^2+C_b^2}{(\lambda _\alpha -\frac i2
)^2+C_b^2}\prod_{l=1}^N\frac{\lambda _\alpha -\frac 12\cot \frac{k_l} 2+
\frac i2}{\lambda _\alpha -\frac 12\cot \frac{k_l}2-\frac i2}\frac{\lambda
_\alpha +\frac 12\cot \frac{k_l}2+\frac i2}{\lambda _\alpha +\frac 12\cot 
\frac{k_l}2-\frac i2} 
\]

\begin{equation}
\qquad \qquad \qquad =\prod_{\beta =1(\beta \neq \alpha )}^M\frac{\lambda
_\alpha -\lambda _\beta +i}{\lambda _\alpha -\lambda _\beta -i}\frac{\lambda
_\alpha +\lambda _\beta +i}{\lambda _\alpha +\lambda _\beta -i}\qquad
(\alpha =1,2,\cdots ,M).
\end{equation}
$R_a(k_j,\widehat{\sigma }_j)$ and $R_b(-k_j,\widehat{\sigma }_j)$ denote
that the boundary matrices at the left and the right ends of the system are
independent on the spin $\sigma _j$ respectively. Notice that the number of
down spins is less than $N+2$ for the system with $N$ conduction electrons.
In the following section, we focus the discussions on the system with the
boundary matrices depending on the spins of the electrons at both ends of
the chain. Set 
\[
\theta _a(k)=\frac 1i\ln \frac{\left( 4C_a^2-3\right) \cos
k-4C_a^2+5+4iC_a\sin k}{\left( 4C_a^2-3\right) \cos k-4C_a^2+5-4iC_a\sin k}, 
\]
\begin{equation}
\theta _b(k)=\frac 1i\ln \frac{\left( 4C_b^2-3\right) \cos
k-4C_b^2+5+4iC_b\sin k}{\left( 4C_b^2-3\right) \cos k-4C_b^2+5-4iC_b\sin k}.
\end{equation}
From relation (\ref{ee6}) and 
\begin{equation}
\exp \left[ i\varphi _b(k_j)\right] =\frac{J_b(\cot \frac{k_j}2+2iC_b)\exp
(ik_j)+i[4+(4V_b-J_b)\exp (ik_j)]}{J_b(\cot \frac{k_j}2-2iC_b)\exp
(-ik_j)-i[4+(4V_b-J_b)\exp (-ik_j)]}
\end{equation}
we get the following expressions. When $J=2$ and $V=-\frac 12,$ we have that 
\begin{equation}
\exp \left[ i\varphi _a(k)\right] =\left\{ 
\begin{array}{c}
\exp (ik)\qquad \qquad \,\text{for }J_a=-\frac 8{\left( 2C_a-1\right) \left(
2C_a+3\right) },V_a=\frac{3-4C_a^2}{\left( 2C_a-1\right) \left(
2C_a+3\right) } \\ 
\exp [ik+i\theta _a(k)]\quad \text{for }J_a=-\frac 8{\left( 2C_a+1\right)
\left( 2C_a-3\right) },V_a=\frac{3-4C_a^2}{\left( 2C_a+1\right) \left(
2C_a-3\right) }
\end{array}
.\right.
\end{equation}
When $J=-2$ and $V=\frac 12,$ we have that 
\begin{equation}
\exp \left[ i\varphi _a(k)\right] =\left\{ 
\begin{array}{c}
-\exp (ik)\qquad \quad \qquad \,\text{for }J_a=\frac 8{\left( 2C_a-1\right)
\left( 2C_a+3\right) },V_a=\frac{4C_a^2-3}{\left( 2C_a-1\right) \left(
2C_a+3\right) } \\ 
-\exp [ik-i\theta _a(\pi -k)]\text{for }J_a=\frac 8{\left( 2C_a+1\right)
\left( 2C_a-3\right) },V_a=\frac{4C_a^2-3}{\left( 2C_a+1\right) \left(
2C_a-3\right) }
\end{array}
.\right.
\end{equation}
When $J=2$ and $V=\frac 32,$ we have that 
\begin{equation}
\exp \left[ i\varphi _a(k)\right] =\left\{ 
\begin{array}{c}
-\exp [ik+i\theta _a(\pi -k)]\text{for }J_a=\frac{-8}{\left( 2C_a-1\right)
\left( 2C_a+3\right) },V_a=\frac{4C_a^2-7}{\left( 2C_a-1\right) \left(
2C_a+3\right) } \\ 
-\exp (ik)\qquad \qquad \quad \,\text{for }J_a=\frac{-8}{\left(
2C_a+1\right) \left( 2C_a-3\right) },V_a=\frac{4C_a^2-7}{\left(
2C_a+1\right) \left( 2C_a-3\right) }
\end{array}
.\right.
\end{equation}
When $J=-2$ and $V=-\frac 32,$ we have that 
\begin{equation}
\exp \left[ i\varphi _a(k)\right] =\left\{ 
\begin{array}{c}
\exp [ik+i\theta _a(-k)]\text{for }J_a=\frac 8{\left( 2C_a-1\right) \left(
2C_a+3\right) },V_a=\frac{7-4C_a^2}{\left( 2C_a-1\right) \left(
2C_a+3\right) } \\ 
\exp (ik)\qquad \qquad \quad \,\text{for }J_a=\frac 8{\left( 2C_a+1\right)
\left( 2C_a-3\right) },V_a=\frac{7-4C_a^2}{\left( 2C_a+1\right) \left(
2C_a-3\right) }
\end{array}
.\right.
\end{equation}
The expressions of $\exp \left[ i\varphi _b(k)\right] $ can be obtained by
substituting the index $a$ of $b$ in the above relations. Then, without loss
any generalization, we can choose that 
\begin{eqnarray}
J_a &=&-\frac 8{\left( 2C_a-1\right) \left( 2C_a+3\right) },V_a=\frac{
3-4C_a^2}{\left( 2C_a-1\right) \left( 2C_a+3\right) },  \nonumber \\
J_b &=&-\frac 8{\left( 2C_b-1\right) \left( 2C_b+3\right) },V_b=\frac{
3-4C_b^2}{\left( 2C_b-1\right) \left( 2C_b+3\right) }
\end{eqnarray}
for $J=2$ and $V=-\frac 12.$ The Bethe Ansatz equations take the forms as 
\begin{equation}
\frac{q_{j+i(C_a+1)}}{q_{j-i(C_a+1)}}\frac{q_{j+i(C_b+1)}}{q_{j-i(C_b+1)}}
\exp (2ik_jG)=\prod_{\beta =1}^M\frac{q_j-\lambda _\beta +\frac i2}{
q_j-\lambda _\beta -\frac i2}\frac{q_j+\lambda _\beta +\frac i2}{q_j+\lambda
_\beta -\frac i2},  \label{e3}
\end{equation}
\[
\frac{(\lambda _\alpha +\frac i2)^2+C_a^2}{(\lambda _\alpha -\frac i2
)^2+C_a^2}\frac{(\lambda _\alpha +\frac i2)^2+C_b^2}{(\lambda _\alpha -\frac 
i2)^2+C_b^2}\prod_{l=1}^N\frac{\lambda _\alpha -q_l+\frac i2}{\lambda
_\alpha -q_l-\frac i2}\frac{\lambda _\alpha +q_l+\frac i2}{\lambda _\alpha
+q_l-\frac i2}\qquad \qquad \qquad \qquad 
\]
\begin{equation}
\qquad =\prod_{\beta =1(\beta \neq \alpha )}^M\frac{\lambda _\alpha -\lambda
_\beta +i}{\lambda _\alpha -\lambda _\beta -i}\frac{\lambda _\alpha +\lambda
_\beta +i}{\lambda _\alpha +\lambda _\beta -i},  \label{e4}
\end{equation}
where $q_j=\frac 12\cot \frac{k_j}2.$ For the case of $J=-2$ and $V=\frac 12
, $ the Bethe Ansatz equations have also the forms (\ref{e3}) and (\ref{e4}
) with $q_j=-\frac 12\tan \frac{k_j}2$ and the Kondo coupling constants
should be 
\begin{eqnarray}
J_a &=&\frac 8{\left( 2C_a-1\right) \left( 2C_a+3\right) },V_a=\frac{
4C_a^2-3 }{\left( 2C_a-1\right) \left( 2C_a+3\right) },  \nonumber \\
J_b &=&\frac 8{\left( 2C_b-1\right) \left( 2C_b+3\right) },V_b=\frac{
4C_b^2-3 }{\left( 2C_b-1\right) \left( 2C_b+3\right) }.
\end{eqnarray}
If $J=2$ and $V=\frac 32,$ we choose that 
\begin{eqnarray}
J_a &=&-\frac 8{\left( 2C_a+1\right) \left( 2C_a-3\right) },V_a=\frac{
4C_a^2-7}{\left( 2C_a+1\right) \left( 2C_a-3\right) },  \nonumber \\
J_b &=&-\frac 8{\left( 2C_b+1\right) \left( 2C_b-3\right) },V_b=\frac{
4C_b^2-7}{\left( 2C_b+1\right) \left( 2C_b-3\right) },
\end{eqnarray}
and the Bethe Ansatz equations are 
\begin{equation}
\exp (2ik_jG)\frac{q_{j+i(C_a+1)}}{q_{j-i(C_a+1)}}\frac{q_{j+i(C_b+1)}}{
q_{j-i(C_b+1)}}\prod_{l=1(l\neq j)}^N\frac{q_j-q_l+i}{q_j-q_l-i}\frac{
q_j+q_l+i}{q_j+q_l-i}\qquad \qquad  \label{eee3}
\end{equation}
\[
\qquad \qquad \qquad \qquad \qquad \qquad =\prod_{\beta =1}^M\frac{
q_j-\lambda _\beta +\frac i2}{q_j-\lambda _\beta -\frac i2}\frac{q_j+\lambda
_\beta +\frac i2}{q_j+\lambda _\beta -\frac i2} 
\]
and relation (\ref{e4}) with $q_j=\frac 12\tan \frac{k_j}2.$ They are also
the Bethe Ansatz equations for $J=-2$ and $V=-\frac 32$ with $q_j=-\frac 12
\cot \frac{k_j}2$ and 
\begin{eqnarray}
J_a &=&\frac 8{\left( 2C_a+1\right) \left( 2C_a-3\right) },V_a=\frac{
7-4C_a^2 }{\left( 2C_a+1\right) \left( 2C_a-3\right) },  \nonumber \\
J_b &=&\frac 8{\left( 2C_b+1\right) \left( 2C_b-3\right) },V_b=\frac{
7-4C_b^2 }{\left( 2C_b+1\right) \left( 2C_b-3\right) }.
\end{eqnarray}

\section{Ground State}

In this paper we restrict the discussions of the properties of ground state
to the case of $J=\pm 2$ and $V=\mp \frac 12.$ The case of $J=2$ and $V=
\frac 32$ were studied in \cite{WDHP}. The eigenvalue of the hamiltonian is 
\begin{equation}
E=\mp 2N\pm \sum_{j=1}^N\frac 1{q_j^2+\frac 14}
\end{equation}
for $J=2$, $V=-\frac 12$ with $q_j=\frac 12\cot \frac{k_j}2$ and $J=-2$, $V=
\frac 12$ with $q_j=-\frac 12\tan \frac{k_j}2$, respectively. They satisfy
the Bethe Ansatz equations (\ref{e3}) and (\ref{e4}) from which the integral
equations are derived.

\subsection{Integral Equations}

Following \cite{401}, we introduce the notation 
\[
e(x)\equiv \frac{x+i}{x-i}. 
\]
Then, from relations (\ref{e3}) and (\ref{e4})we get 
\begin{equation}
e\left( \frac{q_j}{1+C_a}\right) e\left( \frac{q_j}{1+C_b}\right)
e(2q_j)^{2G}=\prod_{\beta =1}^Me\left( 2q_j-2\lambda _\beta \right) e\left(
2q_j+2\lambda _\beta \right) ,  \label{eee8}
\end{equation}
\[
e\left( \frac{\lambda _\alpha }{\frac 12-C_a}\right) e\left( \frac{\lambda
_\alpha }{\frac 12+C_a}\right) e\left( \frac{\lambda _\alpha }{\frac 12-C_b}
\right) e\left( \frac{\lambda _\alpha }{\frac 12+C_b}\right) \qquad \qquad
\qquad 
\]
\begin{equation}
\cdot \prod_{l=1}^Ne\left( 2\lambda _\alpha -2q_l\right) e\left( 2\lambda
_\alpha +2q_l\right) =\prod_{\beta =1(\beta \neq \alpha )}^Me\left( \lambda
_\alpha -\lambda _\beta \right) e\left( \lambda _\alpha +\lambda _\beta
\right) ,  \label{ee9}
\end{equation}
where $j=1,2,\cdots ,N;$ $\alpha =1,2,\cdots ,M$ and $e(\pm \infty )=1.$
Considering that the parameter $q_j$ can take complex values, the general
structure for $\{q_j\}_{j=1,2,...,N}$ should be consisting of $M^{\prime }$
pairs of$\ q_\alpha ^{\pm }=\lambda _\alpha \pm \frac i2+O(\exp (-\delta
G),\alpha =1,\cdots ,M^{\prime }$ and $M^{\prime \prime }$ pairs of$\ 
\widetilde{q}_\alpha ^{\pm }=-\widetilde{\lambda }_\alpha \pm \frac i2
+O(\exp (-\delta G),\widetilde{\lambda }_\alpha \in \{\lambda _\beta \}$ and
remaining $N-2(M^{\prime }+M^{\prime \prime })$ non-pairing $q_j^{\prime }$
s. To be more precise, we use

\[
Q\equiv \{q_j\mid j=1,2,...,N\}=X^{\prime }\cup X^{\prime \prime }\cup Y,
\]
where 
\begin{eqnarray}
X^{\prime } &=&\{q_\alpha ^{\pm }=\lambda _\alpha \pm \frac i2+O(\exp
(-\delta G)\mid \alpha =1,\cdots ,M^{\prime }\},  \nonumber \\
X^{\prime \prime } &=&\{\widetilde{q}_\alpha ^{\pm }=-\widetilde{\lambda }
_\alpha \pm \frac i2+O(\exp (-\delta G)\mid \widetilde{\lambda }_\alpha \in
\{\lambda _\beta \},\alpha =1,\cdots ,M^{\prime \prime }\}, \\
Y &=&Q\backslash (X^{\prime }\cup X^{\prime \prime }).  \nonumber
\end{eqnarray}
Obviously, the non-pairing $q_j$ satisfies equation (\ref{eee8}) with $
j=1,2,\cdots ,N-2(M^{\prime }+M^{\prime \prime })$. When $q_j\in X^{\prime },
$ from equation (\ref{eee8}), we have 
\[
e\left( \frac{\lambda _\alpha }{\frac 32+C_a}\right) e\left( \frac{\lambda
_\alpha }{\frac 12+C_a}\right) e\left( \frac{\lambda _\alpha }{\frac 32+C_b}
\right) e\left( \frac{\lambda _\alpha }{\frac 12+C_b}\right) e\left( \lambda
_\alpha \right) ^{2G}\qquad \qquad \qquad \qquad \qquad 
\]
\begin{equation}
=e(2q_\alpha ^{+}-2\lambda _\alpha )e(2q_\alpha ^{-}-2\lambda _\alpha
)e(2\lambda _\alpha )\prod_{\beta =1,(\beta \neq \alpha )}^Me(\lambda
_\alpha -\lambda _\beta )e(\lambda _\alpha +\lambda _\beta ),\alpha
=1,2,\cdots ,M^{\prime }.  \label{ee10}
\end{equation}
From equation (\ref{ee9}) we have 
\begin{eqnarray*}
&&e\left( \frac{\lambda _\alpha }{\frac 12-C_a}\right) e\left( \frac{\lambda
_\alpha }{\frac 12+C_a}\right) e\left( \frac{\lambda _\alpha }{\frac 12-C_b}
\right) e\left( \frac{\lambda _\alpha }{\frac 12+C_b}\right) e(2\lambda
_\alpha ) \\
&&\cdot \prod_{l=1}^{N-2M_{-}}e(2\lambda _\alpha -2q_l)e(2\lambda _\alpha
+2q_l)\prod_{\beta =1}^{M-M_{-}}e(\widehat{\lambda }_\beta -\lambda _\alpha
)e(-\widehat{\lambda }_\beta -\lambda _\alpha )
\end{eqnarray*}
\begin{equation}
=e(2q_\alpha ^{+}-2\lambda _\alpha )e(2q_\alpha ^{-}-2\lambda _\alpha )
\label{ee1011}
\end{equation}
where $M_{-}=M^{\prime }+M^{\prime \prime }$ and $\widehat{\lambda }_\beta $
( $\beta =1,2,\cdots ,M-M_{-})$ are the parameters describing the down spins
but having no contributions to the bound states. With the help of the above
relation, equation (\ref{ee10}) becomes 
\[
e\left( \frac{\lambda _\alpha }{C_a+\frac 32}\right) e\left( \frac{\lambda
_\alpha }{C_a-\frac 12}\right) e\left( \frac{\lambda _\alpha }{C_b+\frac 32}
\right) e\left( \frac{\lambda _\alpha }{C_b-\frac 12}\right) e\left( \lambda
_\alpha \right) ^{2G}\qquad \qquad \qquad \qquad 
\]
\begin{eqnarray}
\qquad  &=&e\left( 2\lambda _\alpha \right)
^2\prod_{l=1}^{N-2M_{-}}e(2\lambda _\alpha -2q_l)e(2\lambda _\alpha
+2q_l)\prod_{\beta =1(\beta \neq \alpha )}^{M^{\prime }}e(\lambda _\alpha
-\lambda _\beta )e(\lambda _\alpha +\lambda _\beta )  \nonumber \\
&&\cdot \prod_{\beta =1}^{M^{\prime \prime }}e(\lambda _\alpha -\widetilde{
\lambda }_\beta )e(\lambda _\alpha +\widetilde{\lambda }_\beta
),\;\;\;\alpha =1,\cdots ,M^{\prime }.  \label{ee11}
\end{eqnarray}
Similarly, when $q_j\in X^{\prime \prime },$we get the following equation 
\[
e\left( \frac{\widetilde{\lambda }_\alpha }{C_a+\frac 32}\right) e\left( 
\frac{\widetilde{\lambda }_\alpha }{C_a-\frac 12}\right) e\left( \frac{
\widetilde{\lambda }_\alpha }{C_b+\frac 32}\right) e\left( \frac{\widetilde{
\lambda }_\alpha }{C_b-\frac 12}\right) e\left( \widetilde{\lambda }_\alpha
\right) ^{2G}\qquad \qquad \qquad 
\]
\begin{eqnarray}
\qquad  &=&e\left( 2\widetilde{\lambda }_\alpha \right)
^2\prod_{l=1}^{N-2M_{-}}e(2\widetilde{\lambda }_\alpha -2q_l)e(2\widetilde{
\lambda }_\alpha +2q_l)\prod_{\beta =1}^{M^{\prime }}e(\widetilde{\lambda }
_\alpha -\lambda _\beta )e(\widetilde{\lambda }_\alpha +\lambda _\beta ) 
\nonumber \\
&&\cdot \prod_{\beta =1(\beta \neq \alpha )}^{M^{\prime \prime }}e(
\widetilde{\lambda }_\alpha -\widetilde{\lambda }_\beta )e(\widetilde{
\lambda }_\alpha +\widetilde{\lambda }_\beta ),\;\;\;\alpha =1,\cdots
,M^{\prime \prime }  \label{ee12}
\end{eqnarray}
The two equations (ee11) and (ee12) can be combined in a single equation. 
\[
e\left( \frac{\lambda _\alpha }{C_a+\frac 32}\right) e\left( \frac{\lambda
_\alpha }{C_a-\frac 12}\right) e\left( \frac{\lambda _\alpha }{C_b+\frac 32}
\right) e\left( \frac{\lambda _\alpha }{C_b-\frac 12}\right) e\left( \lambda
_\alpha \right) ^{2G}\qquad \qquad \qquad \qquad 
\]
\begin{eqnarray}
\qquad  &=&e\left( 2\lambda _\alpha \right)
^2\prod_{l=1}^{N-2M_{-}}e(2\lambda _\alpha -2q_l)e(2\lambda _\alpha +2q_l) 
\nonumber \\
&&\cdot \prod_{\beta =1(\beta \neq \alpha )}^{M_{-}}e(\lambda _\alpha
-\lambda _\beta )e(\lambda _\alpha +\lambda _\beta ),\;\alpha =1,2,...,M_{-},
\label{ee13}
\end{eqnarray}
with the new $\lambda _\alpha $ defined by 
\[
\lambda _\alpha =\left\{ 
\begin{array}{ccc}
\lambda _\alpha  & \text{when} & \alpha =1,2,...,M^{\prime } \\ 
\widetilde{\lambda }_{M^{\prime }-\alpha } & \text{when} & \alpha =M^{\prime
}+1,M^{\prime }+2,...,M_{-}
\end{array}
\right. 
\]
The parameters $\widehat{\lambda }_\alpha $ $(\alpha =1,2,\cdots ,M-M_{-}),$
in view of (\ref{ee1011}), satisfy 
\[
e\left( \frac{\widehat{\lambda }_\alpha }{C_a+\frac 12}\right) e\left( \frac{
\widehat{\lambda }_\alpha }{C_b+\frac 12}\right)
\prod_{l=1}^{N-2M_{-}}e\left( 2\widehat{\lambda }_\alpha -2q_l\right)
e\left( 2\widehat{\lambda }_\alpha +2q_l\right) 
\]
\begin{equation}
=e\left( \frac{\widehat{\lambda }_\alpha }{C_a-\frac 12}\right) e\left( 
\frac{\widehat{\lambda }_\alpha }{C_b-\frac 12}\right) \prod_{\beta =1(\beta
\neq \alpha )}^{M-M_{-}}e\left( \widehat{\lambda }_\alpha -\widehat{\lambda }
_\beta \right) e\left( \widehat{\lambda }_\alpha +\widehat{\lambda }_\beta
\right) .  \label{ee14}
\end{equation}
The non-pairing $q_j$ (i,e. $q_j\in Y)$satisfies 
\begin{eqnarray}
e\left( \frac{q_j}{C_a+1}\right) e\left( \frac{q_j}{C_b+1}\right)
e(2q_j)^{2G} &=&\prod_{\beta =1}^{M_{-}}e\left( 2q_j-2\lambda _\beta \right)
e\left( 2q_j+2\lambda _\beta \right)   \nonumber \\
&&\cdot \prod_{\beta =1}^{M-M_{-}}e\left( 2q_j-2\widehat{\lambda }_\beta
\right) e\left( 2q_j+2\widehat{\lambda }_\beta \right) ,  \label{ee15}
\end{eqnarray}
where $j=1,2,\cdots ,N-2M_{-}$ and $e(\pm \infty )=1.$ Setting 
\[
\theta (x)\equiv 2\tan ^{-1}x,\qquad -\pi <\theta \leq \pi ,
\]
we have 
\[
e(x)=\exp [i(\pi -\theta (x))].
\]
The logarithms of the equations (\ref{ee13}), (\ref{ee14}) and (\ref{ee15})
give, respectively, 
\[
\theta \left( \frac{\lambda _\alpha }{C_a+\frac 32}\right) +\theta \left( 
\frac{\lambda _\alpha }{C_a-\frac 12}\right) +\theta \left( \frac{\lambda
_\alpha }{C_b+\frac 32}\right) +\theta \left( \frac{\lambda _\alpha }{C_b-
\frac 12}\right) +2G\theta \left( \lambda _\alpha \right) \qquad \qquad
\qquad \ \qquad 
\]
\begin{eqnarray}
\qquad  &=&4\pi J_\alpha +\theta \left( 2\lambda _\alpha \right)
+\sum_{l=1}^{N-2M_{-}}[\theta (2\lambda _\alpha -2q_l)+\theta (2\lambda
_\alpha +2q_l)]  \nonumber \\
&&+\sum_{\beta =1}^{M_{-}}[\theta (\lambda _\alpha -\lambda _\beta )+\theta
(\lambda _\alpha +\lambda _\beta )]  \label{ee16}
\end{eqnarray}
with $\alpha =1,2,\cdots ,M_{-}$ and integers or half-integer $J_\alpha $; 
\[
\theta \left( \frac{\widehat{\lambda }_\alpha }{C_a+\frac 12}\right) +\theta
\left( \frac{\widehat{\lambda }_\alpha }{C_b+\frac 12}\right)
+\sum_{l=1}^{N-2M_{-}}[\theta (2\widehat{\lambda }_\alpha -2q_l)+\theta (2
\widehat{\lambda }_\alpha +2q_l)]\qquad \qquad \qquad 
\]
\begin{eqnarray}
\qquad  &=&4\pi \widehat{J}_\alpha -\theta (2\lambda _\alpha )+\theta \left( 
\frac{\widehat{\lambda }_\alpha }{C_a-\frac 12}\right) +\theta \left( \frac{
\widehat{\lambda }_\alpha }{C_b-\frac 12}\right)   \nonumber \\
&&+\sum_{\beta =1}^{M-M_{-}}[\theta (\widehat{\lambda }_\alpha -\widehat{
\lambda }_\beta )+\theta (\widehat{\lambda }_\alpha +\widehat{\lambda }
_\beta )]  \label{ee17}
\end{eqnarray}
with $\alpha =1,2,\cdots ,M-M_{-}$ and integers or half-integer $\widehat{J}
_\alpha $; 
\[
\theta \left( \frac{q_j}{C_a+1}\right) +\theta \left( \frac{q_j}{C_b+1}
\right) +2G\theta (2q_j)\qquad \ \qquad \qquad \qquad \qquad \qquad \qquad
\qquad 
\]
\begin{equation}
=4\pi I_j+\sum_{\beta =1}^{M_{-}}[\theta (2q_j-2\lambda _\beta )+\theta
(2q_j+2\lambda _\beta )]+\sum_{\beta =1}^{M-M_{-}}[\theta (2q_j-2\widehat{
\lambda }_\beta )+\theta (2q_j+2\widehat{\lambda }_\beta )]  \label{ee18}
\end{equation}
with $j=1,2,\cdots ,N-2M_{-}$and integers or half-integer $I_j$. By setting 
\begin{equation}
\frac d{dx}\theta \left[ k\left( x+c\right) \right] =2\pi a\left( x+c,\frac 1
k\right) ,  \label{ee19}
\end{equation}
the equations (\ref{ee16}), (\ref{ee17}) and (\ref{ee18}) can be changed
into the forms 
\[
a\left( \lambda _\alpha ,C_a+\frac 32\right) +a\left( \lambda _\alpha ,C_a-
\frac 12\right) +a\left( \lambda _\alpha ,C_b+\frac 32\right) +a\left(
\lambda _\alpha ,C_b-\frac 12\right) +2Ga\left( \lambda _\alpha ,1\right) 
\]
\begin{eqnarray}
&=&\frac{dJ_\alpha }{d\lambda _\alpha }+a\left( \lambda _\alpha ,\frac 12
\right) +\sum_{l=1}^{N-2M_{-}}\left[ a\left( \lambda _\alpha -q_l,\frac 12
\right) +a\left( \lambda _\alpha +q_l,\frac 12\right) \right]   \nonumber \\
&&+\sum_{\beta =1}^{M_{-}}\left[ a\left( \lambda _\alpha -\lambda _\beta
,1\right) +a\left( \lambda _\alpha +\lambda _\beta ,1\right) \right] 
\label{ee20}
\end{eqnarray}
with $\alpha =1,2,\cdots ,M_{-};$ 
\[
a\left( \widehat{\lambda }_\alpha ,C_a+\frac 12\right) +a\left( \widehat{
\lambda }_\alpha ,C_b+\frac 12\right) +\sum_{l=1}^{N-2M_{-}}\left[ a\left( 
\widehat{\lambda }_\alpha -q_l,\frac 12\right) +a\left( \widehat{\lambda }
_\alpha +q_l,\frac 12\right) \right] 
\]
\begin{eqnarray}
&=&\frac{d\widehat{J}_\alpha }{d\widehat{\lambda }_\alpha }-a\left( \widehat{
\lambda }_\alpha ,\frac 12\right) +a\left( \widehat{\lambda }_\alpha ,C_a-
\frac 12\right) +a\left( \widehat{\lambda }_\alpha ,C_b-\frac 12\right)  
\nonumber \\
&&+\sum_{\beta =1}^{M-M_{-}}\left[ a\left( \widehat{\lambda }_\alpha -
\widehat{\lambda }_\beta ,1\right) +a\left( \widehat{\lambda }_\alpha +
\widehat{\lambda }_\beta ,1\right) \right]   \label{ee21}
\end{eqnarray}
with $\alpha =1,2,\cdots ,M-M_{-};$ 
\[
a\left( q_j,C_a+1\right) +a\left( q_j,C_b+1\right) +2Ga\left( q_j,\frac 12
\right) \qquad \qquad \qquad \qquad 
\]
\begin{eqnarray}
\qquad \qquad \qquad  &=&\frac{dI_j}{dq_j}+\sum_{\beta =1}^{M_{-}}\left[
a\left( q_j-\lambda _\beta ,\frac 12\right) +a\left( q_j+\lambda _\beta ,
\frac 12\right) \right]   \nonumber \\
&&+\sum_{\beta =1}^{M-M_{-}}\left[ a\left( q_j-\widehat{\lambda }_\beta ,
\frac 12\right) +a\left( q_j+\widehat{\lambda }_\beta ,\frac 12\right)
\right]   \label{ee22}
\end{eqnarray}
with $j=1,2,\cdots ,N-2M_{-}$. We define that 
\begin{eqnarray*}
j(\lambda )\equiv \theta (\lambda )+\qquad \qquad \qquad \qquad \qquad
\qquad \qquad \qquad \qquad \qquad \qquad \qquad \qquad 
\end{eqnarray*}
\begin{eqnarray}
&&+\frac 1{2G}\left\{ \theta \left( \frac \lambda {C_a+\frac 32}\right)
+\theta \left( \frac \lambda {C_a-\frac 12}\right) +\theta \left( \frac 
\lambda {C_b+\frac 32}\right) +\theta \left( \frac \lambda {C_b-\frac 12}
\right) -\theta (2\lambda )\right\}   \nonumber \\
&&-\frac 1{2G}\left\{ \sum_{l=1}^{N-2M_{-}}[\theta (2\lambda -2q_l)+\theta
(2\lambda +2q_l)]+\sum_{\beta =1}^{M_{-}}[\theta (\lambda -\lambda _\beta
)+\theta (\lambda +\lambda _\beta )]\right\} ,
\end{eqnarray}
\[
\widehat{j}(\widehat{\lambda })\equiv \frac 1{2G}\left\{ \theta \left( \frac{
\widehat{\lambda }}{C_a+\frac 12}\right) -\theta \left( \frac{\widehat{
\lambda }}{C_a-\frac 12}\right) +\theta \left( \frac{\widehat{\lambda }}{C_b+
\frac 12}\right) -\theta \left( \frac{\widehat{\lambda }}{C_b-\frac 12}
\right) +\theta (2\widehat{\lambda })\right\} 
\]
\begin{equation}
+\frac 1{2G}\left\{ \sum_{l=1}^{N-2M_{-}}[\theta (2\widehat{\lambda }
-2q_l)+\theta (2\widehat{\lambda }+2q_l)]-\sum_{\beta =1}^{M-M_{-}}[\theta (
\widehat{\lambda }-\widehat{\lambda }_\beta )+\theta (\widehat{\lambda }+
\widehat{\lambda }_\beta )]\right\} ,
\end{equation}
\[
h(q)\equiv \theta (2q)+\frac 1{2G}\left\{ \theta \left( \frac q{C_a+1}
\right) +\theta \left( \frac q{C_b+1}\right) \right\} \qquad \qquad \qquad
\qquad \qquad \quad \qquad 
\]
\begin{equation}
-\frac 1{2G}\left\{ \sum_{\beta =1}^{M_{-}}[\theta (2q-2\lambda _\beta
)+\theta (2q+2\lambda _\beta )]+\sum_{\beta =1}^{M-M_{-}}[\theta (2q-2
\widehat{\lambda }_\beta )+\theta (2q+2\widehat{\lambda }_\beta )]\right\} .
\end{equation}
Then, the holes of $\lambda ,\widehat{\lambda }$ and $q$ are defined as the
solutions of 
\[
Gj(\lambda )=2\pi \times \left( \text{omitted }J\right) ,
\]
\begin{equation}
G\widehat{j}(\widehat{\lambda })=2\pi \times \left( \text{omitted }\widehat{J
}\right) ,
\end{equation}
\[
Gh(q)=2\pi \times \left( \text{omitted }I\right) .
\]
By taking the thermodynamic limits, we introduce the distribution functions 
\[
\left. 
\begin{array}{c}
\lambda \rightarrow \sigma (\lambda ) \\ 
q\rightarrow \varrho (q) \\ 
\widehat{\lambda }\rightarrow \widehat{\sigma }(\widehat{\lambda })
\end{array}
\right\} \stackrel{\text{holes}}{\rightarrow }\left\{ 
\begin{array}{c}
\sigma ^h(\lambda ) \\ 
\varrho ^h(q) \\ 
\widehat{\sigma }^h(\widehat{\lambda })
\end{array}
\right. .
\]
So we have that 
\begin{eqnarray}
\frac{dj(\lambda )}{d\lambda } &=&2\pi \left( \sigma (\lambda )+\sigma
^h(\lambda )\right) ,  \nonumber \\
\frac{dh(q)}{dq} &=&2\pi \left( \rho (q)+\varrho ^h(q)\right) , \\
\frac{d\widehat{j}(\widehat{\lambda })}{d\widehat{\lambda }} &=&2\pi \left( 
\widehat{\sigma }(\widehat{\lambda })+\widehat{\sigma }^h(\widehat{\lambda }
)\right) .  \nonumber
\end{eqnarray}
Therefore, the integral equations can be written down as 
\[
2a(\lambda ,1)+\frac 1G\left[ a\left( \lambda ,C_a+\frac 32\right) +a\left(
\lambda ,C_a-\frac 12\right) +a\left( \lambda ,C_b+\frac 32\right) +a\left(
\lambda ,C_b-\frac 12\right) \right] 
\]
\begin{eqnarray}
&=&\frac 1Ga\left( \lambda ,\frac 12\right) +2\sigma (\lambda )+2\sigma
^h(\lambda )+\int d\lambda ^{\prime }\sigma (\lambda ^{\prime })[a(\lambda
-\lambda ^{\prime },1)+a(\lambda +\lambda ^{\prime },1)]  \nonumber \\
&&+\int dq\rho (q)\left[ a\left( \lambda -q,\frac 12\right) +a\left( \lambda
+q,\frac 12\right) \right] ,  \label{ee24}
\end{eqnarray}
\[
2a\left( q,\frac 12\right) +\frac 1G[a(q,C_a+1)+a(q,C_b+1)]=2\rho (q)+2\rho
^h(q)\qquad 
\]
\begin{eqnarray}
&&+\int d\lambda \sigma (\lambda )\left[ a\left( q-\lambda ,\frac 12\right)
+a\left( q+\lambda ,\frac 12\right) \right]   \nonumber \\
&&+\int d\widehat{\lambda }\widehat{\sigma }(\widehat{\lambda })\left[
a\left( q-\widehat{\lambda },\frac 12\right) +a\left( q+\widehat{\lambda },
\frac 12\right) \right] ,  \label{ee25}
\end{eqnarray}
\[
\frac 1G\left[ a\left( \widehat{\lambda },\frac 12\right) +a\left( \widehat{
\lambda },C_a+\frac 12\right) +a\left( \widehat{\lambda },C_b+\frac 12
\right) -a\left( \widehat{\lambda },C_a-\frac 12\right) -a\left( \widehat{
\lambda },C_b-\frac 12\right) \right] 
\]
\[
+\int dq\rho (q)\left[ a\left( \widehat{\lambda }-q,\frac 12\right) +a\left( 
\widehat{\lambda }+q,\frac 12\right) \right] 
\]
\begin{equation}
\qquad \qquad =2\widehat{\sigma }(\widehat{\lambda })+2\widehat{\sigma }^h(
\widehat{\lambda })+\int d\widehat{\lambda }^{\prime }\widehat{\sigma }(
\widehat{\lambda }^{\prime })[a(\widehat{\lambda }-\widehat{\lambda }
^{\prime },1)+a(\widehat{\lambda }+\widehat{\lambda }^{\prime },1)],
\label{ee26}
\end{equation}
where $a(\lambda ,\eta )\equiv \eta /[\pi (\lambda ^2+\eta ^2)]$ with the
arbitrary parameter $\eta $. The terms with factors $1/G$ in the above three
equations describe the finite-size corrections of the system.

\subsection{Properties of Ground State}

For the system with $N$ electrons, by using the distributed functions $
\sigma (\lambda ),\widehat{\sigma }(\widehat{\lambda })$ and $\rho (q),$ the
particle number and magnetization per unit length are given by 
\[
\frac NG=\int dq\rho (q)+2\int d\lambda \sigma (\lambda ), 
\]
\begin{equation}
\frac{S_z}G=\frac 12\int dq\rho (q)-\int d\lambda \widehat{\sigma }(\lambda
).
\end{equation}
The energies per unit length have the forms as 
\begin{equation}
\frac EG=-\frac{2N}G+2\pi \int dq\rho (q)a\left( q,\frac 12\right) +2\pi
\int d\lambda \sigma (\lambda )a(\lambda ,1)
\end{equation}
for the case of $J=2,$ $V=-1/2$ and 
\begin{equation}
\frac EG=\frac{2N}G-2\pi \int dq\rho (q)a\left( q,\frac 12\right) -2\pi \int
d\lambda \sigma (\lambda )a(\lambda ,1)
\end{equation}
for the case of $J=-2,$ $V=1/2$. The relations (\ref{ee24}), (\ref{ee25})
and (\ref{ee26}) become as 
\[
2a(\lambda ,1)=2\sigma (\lambda )+2\sigma ^h(\lambda )+\int d\lambda
^{\prime }\sigma (\lambda ^{\prime })[a(\lambda -\lambda ^{\prime
},1)+a(\lambda +\lambda ^{\prime },1)] 
\]
\begin{equation}
+\int dq\rho (q)\left[ a\left( \lambda -q,\frac 12\right) +a\left( \lambda
+q,\frac 12\right) \right] ,  \label{ee28}
\end{equation}
\[
2a\left( q,\frac 12\right) =2\rho (q)+2\rho ^h(q)+\int d\lambda \sigma
(\lambda )\left[ a\left( q-\lambda ,\frac 12\right) +a\left( q+\lambda ,
\frac 12\right) \right] 
\]
\begin{equation}
+\int d\widehat{\lambda }\widehat{\sigma }(\widehat{\lambda })\left[ a\left(
q-\widehat{\lambda },\frac 12\right) +a\left( q+\widehat{\lambda },\frac 12
\right) \right] ,  \label{ee29}
\end{equation}
\[
\int dq\rho (q)\left[ a\left( \widehat{\lambda }-q,\frac 12\right) +a\left( 
\widehat{\lambda }+q,\frac 12\right) \right] \qquad \qquad \qquad \qquad
\qquad \qquad \qquad \qquad 
\]
\begin{equation}
=2\widehat{\sigma }(\widehat{\lambda })+2\widehat{\sigma }^h(\widehat{
\lambda })+\int d\widehat{\lambda }^{\prime }\widehat{\sigma }(\widehat{
\lambda }^{\prime })[a(\widehat{\lambda }-\widehat{\lambda }^{\prime },1)+a( 
\widehat{\lambda }+\widehat{\lambda }^{\prime },1)],  \label{ee30}
\end{equation}
if we set $G\rightarrow +\infty .$ By Fourier transformation of equation ( 
\ref{ee28}) we get that 
\[
\int dq\rho (q)+2\int d\lambda \sigma (\lambda )+\int d\lambda \sigma
^h(\lambda )=1, 
\]
which gives that $N/G=1-\int d\lambda \sigma ^h(\lambda ).$ Owing to $\sigma
^h(\lambda )\geq 0,$ we have that $N\leq G,$ which coincides with the
single-occupancy of every site. We assume that there is one particle per
lattice site, that is, $N/G=1.$ Then we have $\sigma ^h(\lambda )=0.$ Now we
consider the case of nonmagnetic $\rho (q)=0.$ The relation (\ref{ee28})
turns into 
\begin{equation}
a(\lambda ,1)=\sigma (\lambda )+\int d\lambda ^{\prime }\sigma (\lambda
^{\prime })a(\lambda -\lambda ^{\prime },1).  \label{ee31}
\end{equation}
By Fourier transformation of equation (\ref{ee30}) we have that $\widehat{
\sigma }(\widehat{\lambda })=\widehat{\sigma }^h(\widehat{\lambda })=0,$
which means that $S_z/G=0$ and the system is nonmagnetic . From the above
relation, we have that 
\begin{equation}
\sigma (\lambda )=\frac 1{2\pi }\int_{-\infty }^{+\infty }e^{-i\omega
\lambda }\frac{e^{-\frac{\left| \omega \right| }2}}{2\cosh \frac \omega 2}
d\omega .
\end{equation}
The interesting thing is that the above expression is exactly same as the
integrable narrow-band model with periodic boundary obtained by Schlottmann 
\cite{Sch}. In this way, relation (\ref{ee29}) reduces to 
\[
a\left( q,\frac 12\right) =\rho ^h(q)+\int d\lambda \sigma (\lambda )a\left(
q-\lambda ,\frac 12\right) , 
\]
and it gives that 
\begin{equation}
\rho ^h(q)=\frac 1{2\pi }\int_{-\infty }^{+\infty }\frac{e^{-i\omega q}}{
2\cosh \frac \omega 2}d\omega =\left\{ 
\begin{array}{c}
\frac 12\sec h\left| \pi q\right| ,\text{ for }q\neq 0 \\ 
\frac 12,\text{ for }q=0
\end{array}
\right. .
\end{equation}
The number $M$ of the down spins is equal to $G/2.$ The ground-state energy
is $E/G=-2\ln 2$ for $J=2,$ $V=-1/2,$ which has the same value as the one in
the periodic boundary condition \cite{Sch}. It is due to that the impurities
located at the both ends cause only the finite-size correction of the ground
state energy. For the case of $J=-2,$ $V=1/2,$ corresponding to the
ferrimagnetic state, we have that 
\[
\rho (q)=\frac 1\pi \frac{\frac 12}{q^2+\frac 14}, 
\]
\begin{equation}
\widehat{\sigma }^h(\lambda )=\frac 1\pi \frac 1{\lambda ^2+1},\ \qquad 
\widehat{\sigma }(\lambda )=0,
\end{equation}
by taking into account of $\sigma (\lambda )=\sigma ^h(\lambda )=0.$ Then we
have $E/G=0.$

\section{Finite-Size Correction of the Ground State}

We assume that the distribution functions $\sigma (\lambda ),$ $\rho (q)$
and $\widehat{\sigma }(\widehat{\lambda })$ are even functions about
parameters $\lambda ,$ $q$ and $\widehat{\lambda },$ respectively. Then we
have the following equations: 
\[
a(\lambda ,1)+\frac 1{2G}\left[ a\left( \lambda ,C_a+\frac 32\right)
+a\left( \lambda ,C_a-\frac 12\right) +a\left( \lambda ,C_b+\frac 32\right)
+a\left( \lambda ,C_b-\frac 12\right) \right] 
\]
\begin{equation}
=\frac 1{2G}a\left( \lambda ,\frac 12\right) +\sigma (\lambda )+\sigma
^h(\lambda )+\int d\lambda ^{\prime }\sigma (\lambda ^{\prime })a(\lambda
-\lambda ^{\prime },1)+\int dq\rho (q)a\left( \lambda -q,\frac 12\right) ,
\label{ee51}
\end{equation}
\[
a\left( q,\frac 12\right) +\frac 1{2G}[a(q,C_a+1)+a(q,C_b+1)]\qquad \qquad
\qquad \qquad \qquad \qquad \qquad 
\]
\begin{equation}
=\rho (q)+\rho ^h(q)+\int d\lambda \sigma (\lambda )a\left( q-\lambda ,\frac 
12\right) +\int d\widehat{\lambda }\widehat{\sigma }(\widehat{\lambda }
)a\left( q-\widehat{\lambda },\frac 12\right) ,  \label{ee52}
\end{equation}
\[
\frac 1{2G}\left[ a\left( \widehat{\lambda },\frac 12\right) +a\left( 
\widehat{\lambda },C_a+\frac 12\right) +a\left( \widehat{\lambda },C_b+\frac 
12\right) -a\left( \widehat{\lambda },C_a-\frac 12\right) -a\left( \widehat{
\lambda },C_b-\frac 12\right) \right] 
\]
\[
+\int dq\rho (q)a\left( \widehat{\lambda }-q,\frac 12\right) 
\]
\begin{equation}
\qquad \qquad =\widehat{\sigma }(\widehat{\lambda })+\widehat{\sigma }^h(
\widehat{\lambda })+\int d\widehat{\lambda }^{\prime }\widehat{\sigma }(
\widehat{\lambda }^{\prime })a(\widehat{\lambda }-\widehat{\lambda }^{\prime
},1),  \label{ee53}
\end{equation}
from equations (\ref{ee24}), (\ref{ee25}) and (\ref{ee26}), where $a(\lambda
,\eta )\equiv \eta /[\pi (\lambda ^2+\eta ^2)]$ with the arbitrary real
parameter $\eta $. The terms with factors $1/G$ in the above three equations
describe the finite-size corrections of the system. The energies of the
system can be described by 
\begin{equation}
\frac EG=\mp \frac{2N}G\pm 2\pi \left[ \int dq\rho (q)a\left( q,\frac 12
\right) +\int d\lambda \sigma (\lambda )a\left( \lambda ,1\right) \right]
\end{equation}
for $J=\pm 2,$ $V=\mp 1/2,$ respectively. Setting 
\begin{equation}
S_\eta \equiv sign(n)=\left\{ 
\begin{array}{c}
1,\qquad \eta >0 \\ 
-1,\qquad \eta <0 \\ 
0,\qquad \eta =0
\end{array}
,\right.
\end{equation}
we have that 
\begin{equation}
\widetilde{a}(\omega ,\eta )=S_\eta \exp (-\left| \omega \eta \right| ).
\end{equation}
By Fourier transformation of equation (\ref{ee51}), we have 
\[
\widetilde{\sigma }^h(0)=\frac 1{2G}\left[ S_{C_a+\frac 32}+S_{C_a-\frac 12
}+S_{C_b+\frac 32}+S_{C_b-\frac 12}-1\right] 
\]
for $N/G=1.$ By letting 
\begin{equation}
b(\lambda )=a\left( \lambda ,C_a+\frac 32\right) +a\left( \lambda ,C_a-\frac 
12\right) +a\left( \lambda ,C_b+\frac 32\right) +a\left( \lambda ,C_b-\frac 1
2\right) -a\left( \lambda ,\frac 12\right) ,
\end{equation}
we have that 
\[
\int dq\rho (q)a\left( q,\frac 12\right) +\int d\lambda \sigma (\lambda
)a\left( \lambda ,1\right) =a(0,1)+\frac{b(0)}{2G}-\sigma (0)-\sigma ^h(0). 
\]
We set $\sigma ^h(\lambda )$ $\equiv 0.$ The Kondo coupling constants $C_a$
and $C_b$ should be in the ranges (i) $C_a>1/2,$ $C_b=-3/2$; (ii) $C_a=1/2$, 
$1/2>C_b>-3/2;$ (iii) $1/2>C_a>-3/2,$ $C_b=1/2;$ (iv) $C_a=-3/2,$ $C_b>1/2.$
For the case of $J=2,$ $V=-1/2,$ the ground state energy can be written down
as the form 
\begin{equation}
\frac EG=\frac \pi Gb(0)-2\pi \sigma (0),
\end{equation}
and $\sigma (0)$ should take its largest value. Then we set $\rho (q)=0$ and
obtain that 
\begin{equation}
2\pi \sigma (0)=2\ln 2+\frac 1{2G}\int_{-\infty }^{+\infty }\frac{\widetilde{
b}(\omega )}{1+\exp (-\left| \omega \right| )}d\omega  \label{ee55}
\end{equation}
where 
\[
\widetilde{b}(\omega )=S_{C_a+\frac 32}\exp \left[ -\left| \omega \left( C_a+
\frac 32\right) \right| \right] +S_{C_a-\frac 12}\exp \left[ -\left| \omega
\left( C_a-\frac 12\right) \right| \right] \qquad \qquad \qquad \qquad 
\]
\[
+S_{C_b+\frac 32}\exp \left[ -\left| \omega \left( C_a+\frac 32\right)
\right| \right] +S_{C_b-\frac 12}\exp \left[ -\left| \omega \left( C_b-\frac 
12\right) \right| \right] -\exp \left( -\left| \frac \omega 2\right| \right)
. 
\]
Therefore, the finite-size correction of the ground-state energy due to
impurities is 
\begin{equation}
E^{\prime }=\frac{8(2C_a^2+3C_a+2)}{(2C_a+3)(4C_a^2-1)}-\frac 32-\ln 2+\frac 
\pi 2-2\beta (\frac{2C_a-1}2)
\end{equation}
when $C_a>1/2$ and $C_b=-3/2$, where $\beta $ is defined by $\beta
(x)=\allowbreak \frac 12\left[ \psi \left( \frac{x+1}2\right) -\psi \left( 
\frac x2\right) \right] $ and $\psi (x)=\frac d{dx}\ln \Gamma (x).$ By
taking account of $C_a>1/2,$ then $J_a<0,$ we have $\widehat{\sigma }(
\widehat{\lambda })=0$ for the ground state. From relations (\ref{ee51}), ( 
\ref{ee52}) and (\ref{ee53}), we get that 
\begin{equation}
\widehat{\sigma }^h(\widehat{\lambda })=\frac 1{2G}\left[ a\left( \widehat{
\lambda },\frac 12\right) +a\left( \widehat{\lambda },C_a+\frac 12\right)
+a\left( \widehat{\lambda },-1\right) -a\left( \widehat{\lambda },C_a-\frac 1
2\right) -a\left( \widehat{\lambda },-2\right) \right] ,
\end{equation}
\begin{equation}
\rho ^h(0)=\frac 12+\frac 1{2\pi G}\left[ \frac 1{C_a}-\frac{10}3+\ln 2+
\frac \pi 2-2\beta (C_a)\right] ,
\end{equation}
\begin{equation}
\rho ^h(q)=\frac 12\sec h\left| \pi q\right| +\frac 1{2G}\left[
a(q,C_a+1)-a\left( q,\frac 12\right) \right] -\frac 1{4\pi G}\int_0^{+\infty
}\frac{\widetilde{b}(\omega )\cos (\omega q)}{\cosh \frac \omega 2}d\omega
\end{equation}
for $q\neq 0.$ When $C_a=1/2,$ $-3/2<C_b<1/2,$ from relation (\ref{ee55}),
we have that 
\begin{equation}
2\pi \sigma (0)=2\ln 2+\frac 1{2G}\left\{ 2(1-\ln 2)-\pi +2\left[ \beta
\left( \frac{2C_b+3}2\right) -\beta \left( \frac{1-2C_b}2\right) \right]
\right\} .
\end{equation}
Then, the finite-size correction of the ground state has the form 
\begin{eqnarray}
E^{\prime } &=&\frac \pi 2\left[ \cot \left( \frac 14\pi +\frac 12\pi
C_b\right) +\tan \left( \frac 14\pi +\frac 12\pi C_b\right) +1\right] 
\nonumber \\
&&+\ln 2-\frac 12\frac{44C_b^2-26C_b-35+40C_b^3}{\left( 2C_b+3\right) \left(
2C_b-1\right) \left( 2C_b+1\right) }.
\end{eqnarray}
By taking account of $J_b>0,$ we have $\widehat{\sigma }^h(\widehat{\lambda }
)=0.$ From relations (\ref{ee51}), (\ref{ee52}) and (\ref{ee53}), we obtain
that 
\begin{eqnarray}
\widehat{\sigma }(\widehat{\lambda }) &=&\frac 1{4\pi G}\int_0^{+\infty
}d\omega \frac{\cos (\omega \lambda )}{\cosh \frac \omega 2}\left\{ 1+\exp
\left( -\left| \frac \omega 2\right| \right) +\right.  \nonumber \\
&&\left. +S_{C_b+\frac 12}\exp \left[ \left| \frac \omega 2\right| -\left|
\omega \left( C_b+\frac 12\right) \right| \right] \right\} ,
\end{eqnarray}
\[
\rho ^h(q)=\frac 12\sec h\left| \pi q\right| +\frac 1{2G}\left[ a\left( q,
\frac 32\right) +a(q,C_b+1)\right] \qquad \qquad \qquad \qquad 
\]
\begin{eqnarray}
&&-\frac 1{4\pi G}\int_0^{+\infty }d\omega \frac{\cos (\omega q)}{\cosh 
\frac \omega 2}\left\{ \exp \left( -\left| 2\omega \right| \right) +\exp
\left( -\left| \omega \right| \right) +\exp \left[ -\left| \omega \left( C_b+
\frac 32\right) \right| \right] \right.  \nonumber \\
&&\left. +S_{C_b+\frac 12}\exp \left[ -\left| \omega \left( C_b+\frac 12
\right) \right| \right] \right\}
\end{eqnarray}
for $q\neq 0$ and 
\begin{equation}
\rho ^h(0)=\left\{ 
\begin{array}{c}
\frac 12\text{ \qquad \qquad \qquad for }C_b=-1 \\ 
\frac 12+\frac 1{2\pi G}\left\{ \frac 1{C_b+1}+\beta (C_b+2)+S_{C_b-\frac 12
}\beta \left( \frac{\left| 2C_b+1\right| +1}2\right) \right\} \text{ for }
C_b\neq -1
\end{array}
\right. .
\end{equation}
The cases of $1/2>C_a>-3/2,$ $C_b=1/2;$ $C_a=-3/2,$ $C_b>1/2$ have the
similar expressions. When $J=-2,$ $V=1/2,$ by similar discussions, the
finite-size correction of the ground-state energy can be written down as 
\begin{equation}
E^{\prime }=\frac 52-\frac{4(2C_a+1)}{(2C_a-1)(2C_a+3)}
\end{equation}
for $C_a>1/2,$ $C_b=-3/2$ and 
\begin{equation}
E^{\prime }=\frac 32-\frac{4(2C_b+1)}{(2C_b-1)(2C_b+3)}
\end{equation}
for $C_a=1/2$, $1/2>C_b>-3/2.$

As the conclusions, an integrable model in one dimension is constructed from 
$t-J$ model where two magnetic impurities are coupled to the system. It
describe the behavior of the strong correlation electrons with Kondo
problem. The spectrums of the system is not linear. The boundary $R$ matrix
depends on the spin and rapidity of the particle and satisfies the
reflecting factorizable condition. The Hamiltonian of the model is
diagonalized exactly by the Bethe-Ansatz method. The integral equations are
derived with the complex ``rapidities'' $q$ which describe the bound states
in the system. The properties of the ground state are discussed and the
finite-size corrections of the ground-state energies are obtained due to the
couplings of the magnetic impurities.

\end{document}